\documentclass[fleqn,usenatbib]{mnras}

\usepackage{newtxtext,newtxmath}

\usepackage[T1]{fontenc}
\usepackage{orcidlink}
\usepackage{caption}
\usepackage{graphicx}
\usepackage{subcaption}
\usepackage{float}
\usepackage{xcolor}
\usepackage{hyperref}

\DeclareRobustCommand{\VAN}[3]{#2}
\let\VANthebibliography\thebibliography
\def\thebibliography{\DeclareRobustCommand{\VAN}[3]{##3}\VANthebibliography}

\usepackage{graphicx}	
\usepackage{amsmath}	

\title[Super-Earth formation in an icy dead zone]{Formation of super-Earths around low-mass stars: evolution of an icy dead zone }

\author[D. A. Arturo Rodriguez \& R. G. Martin]{
Danilo A. Arturo Rodriguez $^{1,2}$\orcidlink{0009-0008-3058-4277} and Rebecca G. Martin$^{1,2}$\orcidlink{0000-0003-2401-7168} 
\\
$^1$Nevada Center for Astrophysics, University of Nevada, Las Vegas 4505 South Maryland Parkway, Las Vegas, NV 89154, USA\\
$^{2}$Department of Physics and Astronomy, University of Nevada, Las Vegas 4505 South Maryland Parkway, Las Vegas, NV 89154, USA\\
}

\date{Accepted XXX. Received YYY; in original form ZZZ}

\pubyear{2025}

\begin{document}
\label{firstpage}
\pagerange{\pageref{firstpage}--\pageref{lastpage}}
\maketitle

\begin{abstract}
Exoplanet observations show that close-in super-Earths are more common around M-dwarfs than around solar mass stars. Since the snow line in a protoplanetary disc plays a crucial role in determining the amount of solid material available for planet formation, we explore the icy regions of protoplanetary discs around stars with masses $0.1$, $0.5$ and $1\rm \, M_\odot$.
In a protoplanetary disc, a dead zone, where the magneto-rotational instability (MRI) is suppressed, provides a quiescent region for solids to settle to the mid-plane and planets to form. Viscosity may be driven in the dead zone by gravitational instability if enough material builds up. Heating from the gravitational instability can trigger the MRI and an accretion outburst onto the star. There may be two icy regions in a disc: (1) far from the star and (2) in the dead zone close to the star. We solve the 1D disc equations to find steady state solutions and time-dependent evolution with different values for the critical surface density in the MRI-active surface layers. Larger surface density in the MRI-active surface layers reduces the extent and lifetime of the inner icy region. The inner icy region in the dead zone around a solar mass star is small and short-lived. Around M-dwarfs, the size of the inner icy region is more persistent and oscillates between the accretion outbursts in the region $0.1-1\,\rm au$. An extended icy region within the dead zone of a disc around M-dwarfs may promote the formation of more numerous and massive close-in super-Earths.
\end{abstract}

\begin{keywords}
accretion, accretion discs -- protoplanetary discs -- planets and satellites: formation -- stars: pre-main-sequence
\end{keywords}



\section{Introduction}
  
Understanding the differences in planetary system architectures around stars of varying masses provides valuable insights into the process of planet formation. It might be expected that higher-mass stars have more massive protoplanetary discs, leading to the formation of larger and more numerous planets. Indeed, this trend is observed for giant planets. The likelihood of a star hosting a giant planet increases with its mass (e.g. \cite{Johnson07, Johnson10}). Giant planets are found around less than 10\% of G-type stars, with this percentage declining further for lower-mass stars \citep{Wittenmyer16, Wittenmyer20}. Occurrence rates for hot Jupiters also vary by stellar type. Only about 0.34\% of M-dwarfs are observed to host hot Jupiters \citep{Obermeier16}. In contrast, FGK stars have a slightly higher occurrence rate of approximately 1\% \citep{Johnson07,Beleznay22}.
However, the occurrence rate of super-Earths with orbital periods shorter than 100 days shows the opposite trend to the giant planets. The occurrence rate is more than three times higher around M-dwarfs compared to F-type stars \citep{Howard12, Dressing12, Mulders15a, Mulders15b, Mulders18, Mulders21, Sabotta21}. It is important to note that these observed trends are influenced by various detection biases inherent to different survey methods. Transit surveys exhibit strong biases toward larger planets with shorter orbital periods, as these produce deeper and more frequent transit signals \citep{Kipping16}. Similarly, radial velocity surveys are more sensitive to massive planets orbiting close to their host stars due to the larger amplitude signals they produce \citep{Cumming08}. These biases are particularly relevant for M-dwarf surveys, where the favorable mass ratios and smaller stellar radii can enhance detection probabilities for planets, potentially enhancing the observed excess of close-in super-Earths \citep{Gaidos24}. Nevertheless, the high occurrence rate of close-in super-Earths around M-dwarfs persists across multiple detection methods and survey campaigns, indicating that it reflects an intrinsic feature of planetary system architectures rather than an artifact of detection biases \citep{Mulders15a, Mulders15b,Sabotta21, Nielsen25}. The super-Earth trend challenges expectations based on protoplanetary disc masses, raising questions about the processes shaping planet formation in these systems. 

Several mechanisms have been proposed to explain the formation of close-in super-Earths, although most require formation farther out followed by inward migration \citep{Mcneil10,Haghighipour13,Inamdar15,Schlichting18, Zawadzki21}. One possible explanation for the excess of super-Earths around low-mass stars is that pebble accretion in the
inner regions is shut off when a giant planet forms and opens a gap in the disc \citep{Lambrechts19,Mulders211}. This suppresses the growth of close-in planets. However, most M-dwarfs do not have giant planets \citep{Child22},  and some studies suggest that super-Earths may be present even with cold giant planets \citep{Barbato18, Zhuw18, Bryan19}. An alternative scenario proposes that fragmentation of a massive disc around an M-dwarf may be possible,
but it requires a disc with a mass of around 30\% of the star \citep{Boss06, Backus16, Mercer20, Haworth20}.

In this work, we consider the protoplanetary disc conditions that may allow for the in situ planet formation of super-Earths. A critical factor affecting the planet formation process is the protoplanetary disc temperature. The availability of solid material for planet formation is approximately four times higher in icy regions where the mid-plane temperature of the disc, $T_{\rm c}$, is lower than the snow line temperature, $T_{\rm snow}$ \citep{Pollack96, Kennedy08}. The snow line temperature is where water transitions to a solid form, occurring at temperatures between 145 and 170$\,$K \citep{Hayashi,Podolak04}. The radius of the snow line represents the distance from the star where the temperature crosses the snow line temperature, $T_{\rm snow}$. 
Observations indicate that the locations of giant planets tend to peak near the snow line for all stellar types \citep{Fernandes19,Child22}, highlighting their importance in shaping planetary system architectures.

In this work, we are interested in comparing the evolution of the icy regions of protoplanetary discs around stars with different masses. The evolution of protoplanetary discs depends upon the rate at which gas can shed its angular momentum and flow inwards. This is determined by the effective viscosity, $\nu$, that is given by the standard $\alpha$-prescription as
\begin{equation}
\nu = \alpha \frac{c_{\rm s}^2}{\Omega},
\end{equation}
\citep{Shakura1973,Pringle1981}, where $c_{\rm s}$ is the local sound speed, $\Omega=\sqrt{G M/R^3}$ is the Keplerian angular velocity, $G$ is the gravitational constant, $M$ is the mass of the star and $R$ is the radius in the disc. The value for $\alpha$ depends upon the driver of the viscosity. Two main instabilities can transport angular momentum: gravitational instability (GI) \citep{Pacz, Rice04, Kratter16} and the magneto-rotational instability (MRI) \citep{Balbus91}.
In the inner disc where the gas is sufficiently hot, the gas is ionized and well coupled to the magnetic field so that the MRI operates \citep{Armitage2001}. At larger radii, where the central temperature is lower, the MRI is suppressed due to the low ionization, and this leads to the formation of a layered disc structure where turbulence occurs only in the thin surface layers that are ionized by cosmic rays or X-rays from the central star. The cold disc mid-plane forms what is known as a "dead zone" \citep{Gammie96, Delage22, Zhu10, Bai11}. In the outer parts of the disc,  at a radius where the surface density is small enough for cosmic rays to penetrate to the mid-plane, the disc becomes fully MRI-active \citep{Glassgold2004}.

In a layered region containing a dead zone, steady-state accretion is not possible, and the dead zone gains mass from the accretion flow near the disc surfaces. As mass builds up in the dead zone, it can become self-gravitating, causing GI. This produces turbulent heating, which raises the ionization level of the gas. This increased ionization may then trigger the MRI, leading to a higher level of disc turbulence and accretion, an outburst \citep{Armitage2001,Martin11, Martin12a, Martin12b,Zhu09,Zhu10b, Zhu10, Bae2013,Zhu14}.  After an outburst, the remaining disc gas cools, the dead zone reforms, and the cycle repeats. In the later stages of the disc evolution, the infall accretion rate decays exponentially, and there is not enough mass flowing through the disc for GI to transport material inward and trigger the MRI, however, there may still be a dead zone present \citep{Armitage2001}. 

The dead zone provides a quiescent region in the disc where solids can settle to the disc mid-plane \citep[e.g.][]{Youdin2002,Youdin2005,Zsom2011} and allow planet formation to proceed \citep{Bai2010,Yang2018}. Gravitational instability may form large bodies in a dense dust layer \citep{Goldreich1973}, but even low levels of turbulence can disrupt this \citep{Dubrulle1995,Cuzzi2008}. Turbulence can also cause destructive collisions between solid bodies, preventing growth \citep{Ida2008}. 
 In a fully ionized disc, high levels are turbulence are expected, corresponding to a \cite{Shakura1973} viscosity $\alpha\approx 0.1-0.3$ \citep{King2007,Martin2019}. However, the observed sizes of protoplanetary discs \citep{Hartmann1998} and of FU Ori \citep{Zhu2007} suggest that turbulence is of the order of $\alpha \approx 0.01$. 

\cite{Vallet23} used steady-state disc models and proposed a solution to the observed excess of close-in super-Earths around low-mass stars. They suggested that extended dead zones may contain an inner icy region, where temperatures remain below the snow line, facilitating the condensation and retention of icy solids. They suggested that the extent of this region is larger in discs around low-mass stars, providing an increased reservoir of solid material for planet formation. This unique structure may support the in-situ formation of super-Earths, explaining the higher occurrence rate around low-mass stars.

In this paper, we first extend the results of \cite{Vallet23} by incorporating an MRI-active layer in the surface of the disc with critical surface density $\Sigma_{\rm crit}$ and exploring conditions under which an icy dead zone can form. In section \ref{sec:eqdisc} we describe the equations for the disc model. In section \ref{sec:result}, we consider fully turbulent steady-state disc solutions and compare those to a disc with a dead zone. We demonstrate that the inner icy region within the dead zone is smaller around a solar-mass star than around an M-dwarf. Additionally, we further show that the size of the inner icy region decreases as the critical surface density increases.
Then, in Section \ref{sec:timedepen}, with time-dependent 1D simulations, we examine the evolution of layered discs and consider how long the icy dead zone is present.
We show how the accretion rate onto the star and the icy regions vary in time. Finally, in Section \ref{conc}, we summarize our results and discuss their implications for the formation of super-Earths.

\section{Time-dependent disc evolution equations}
\label{sec:eqdisc}
In this Section, we first explain the structure of the layered disc and then we describe the 1D time-dependent disc evolution equations. 
M-dwarfs span a broad mass range and so for each disc model we consider three stellar masses, a low-mass M-dwarf ($M=0.1\,\rm M_{\odot}$), a high mass M-dwarf ($M = 0.5\,\rm M_{\odot}$), and a solar mass star ($M=1\,\rm M_\odot$) for comparison. The parameters that we take for the star and the disc are summarized in Table~\ref{tab:params}.

\begin{table*}
\centering
\small
\begin{tabular}{|l|p{7.0cm}|p{6.0cm}|}
\hline
\multicolumn{3}{|c|}{\textbf{Fixed parameters}} \\
\hline
Parameter & Description & Adopted value(s) / note \\
\hline
$M$ & Stellar mass & $1.0\, \rm M_\odot,\;0.5\,M_\odot,\;0.1\,M_\odot$ \\
$R_*$ (pre-MS) & Stellar radius (pre-main sequence) & $3\,R_\odot$ (for $1.0\, \rm M_\odot$), $0.8\,R_\odot$ (for $0.5\, \rm M_\odot$), $0.4\,R_\odot$ (for $0.1\,\rm M_\odot$) \\
$T_*$ (pre-MS) & Stellar effective temperature (pre-main sequence) & $4000\,$K (for $1.0\, \rm M_\odot$), $3500\,$K (for $0.5\, \rm M_\odot$), $2850\,$K (for $0.1\, \rm M_\odot$) \\
$R_{\rm in}$ & Inner disc radius & $5\,R_\odot$ \\
$R_{\rm out}$ & Outer disc radius & $40\,$au \\
$R_{\rm add}$ & Mass addition (infall) radius & $35\,$au \\
$\dot{M}_{\rm i}$ & Initial infall accretion rate & $2\times10^{-5}\, \rm M_\odot\,{\rm yr}^{-1}$ (for all stellar masses) \hspace{3cm} $2\times10^{-6}\, \rm M_\odot\,{\rm yr}^{-1}$ (for the $0.1\, \rm M_\odot$ case only) \\
$t_0$ & Duration of constant infall before infall decay & $1\,$Myr  \\
$t_{\rm ff}$ & Free-fall timescale  & $10^{5}\,$yr \\
$\rho_{\rm cloud}$ & Cloud density used to compute $t_{\rm ff}$ & $4.41\times10^{-19}\,\rm g\,cm^{-3}$  \\
$\Sigma_{\rm crit}$ & Critical surface density for MRI-active surface layers & $2,\;20,\;200\ \rm g\,cm^{-2}$  \\
$T_{\rm crit}$ & Critical temperature for thermal ionization / MRI activation & $800\,$K  \\
$\alpha_{\rm m}$ & MRI-active layer viscosity parameter & $0.01$\\
$Q_{\rm crit}$ & Toomre critical parameter & $2$ \\
$\mu$ & Mean molecular weight & $2.3$ \\
$T_{\rm snow}$ & Snow-line (water ice) temperature & $145\,$K \\
\hline
\end{tabular}

\vspace{0.5cm}

\begin{tabular}{|l|p{13cm}|}
\hline
\multicolumn{2}{|c|}{\textbf{Computed parameters}} \\
\hline
Parameter & Description \\
\hline
$\Omega$ & Keplerian angular frequency \\
$\Sigma$ & Total disc surface density  \\
$\Sigma_{\rm m}$ & MRI-active  layer surface density \\
$\Sigma_{\rm g}$ & Mid-plane surface density \\
$T_{\rm c}$ & Mid-plane temperature  \\
$T_{\rm m}$ & Temperature in MRI-active layers \\
$T_{\rm e}$ & Surface temperature \\
$\nu_{\rm m}$ & Viscosity in MRI-active layers \\
$\nu_{\rm g}$ & Viscosity due to self-gravity (when $Q<Q_{\rm crit}$) \\
$\alpha_{\rm g}$ & Effective viscosity parameter due to self-gravity \\
$c_{\rm m}$ & Sound speed in MRI-active layers \\
$c_{\rm g}$ & Sound speed at the mid-plane\\
$Q$ & Toomre parameter \\
$c_{\rm p}$ & Disc specific heat  \\
$\tau_{\rm m}$ & Optical depth of MRI-active layers \\
$\tau_{\rm g}$ & Optical depth of the mid-plane layer\\
$\tau$ & Total optical depth \\
$\kappa$ & Opacity \\
$Q_-$ & Cooling function \\
$Q_{+}$ & Local heating owing to viscous dissipation \\
$F_{\rm irr}$ & Stellar irradiation flux \\
$\alpha_{\rm irr}$ & Stellar irradiation geometry parameter \\
$T_{\rm irr}$ & Irradiation temperature \\
\hline
\end{tabular}
\caption{List of parameters. Top block: fixed adopted parameters used for the star and the disc. Bottom block: computed parameters (quantities that vary with $R$ and $t$).}
\label{tab:params}
\end{table*}

\subsection{Disc Structure}

 \begin{figure*}
    \centering
    \includegraphics[width=0.7\textwidth]{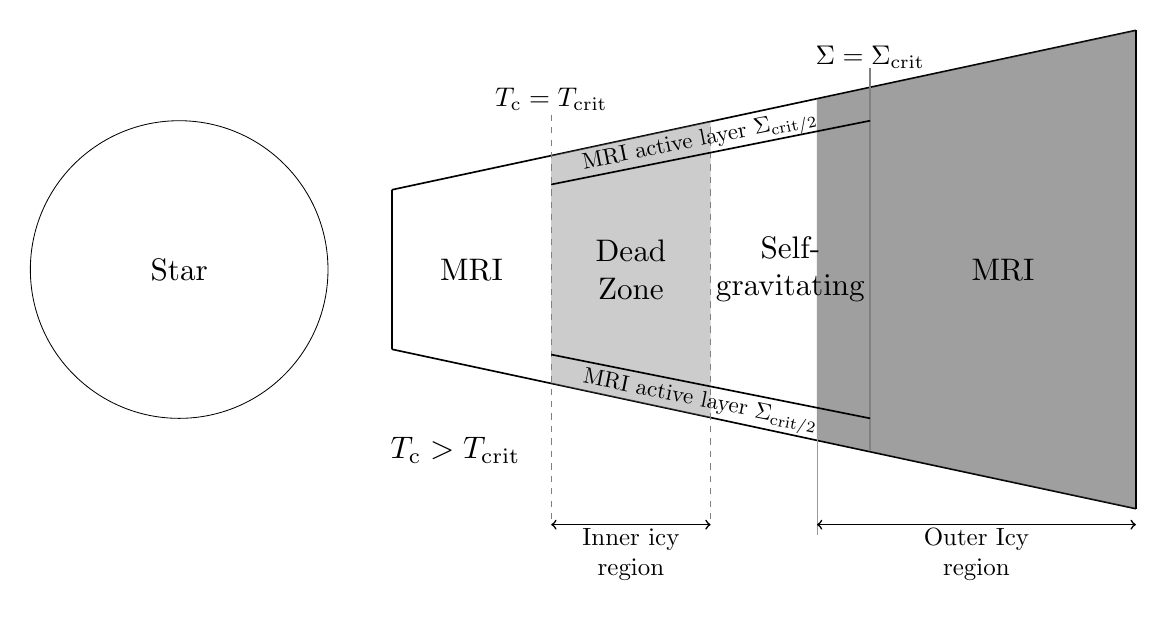}
    \caption{Sketch of the protoplanetary disc structure. In the inner region of the disc, where the gas is sufficiently hot ($T_{\rm c} > T_{\rm crit}$), the gas is ionized and well-coupled to the magnetic field, allowing the MRI to operate effectively. At larger radii, where the temperature is lower, the MRI is suppressed due to insufficient ionization, leading to the formation of a layered disc structure. MRI turbulence occurs  in the thin surface layers ionized by cosmic rays or X-rays (with $\Sigma=\Sigma_{
    \rm crit}$), while the cold mid-plane forms a ``dead zone'' where the MRI is inactive. In the outer parts of the disc, where the surface density is low enough ($\Sigma < \Sigma_{\rm crit}$) for cosmic rays to penetrate to the mid-plane, the disc becomes fully MRI-active again. The outer parts of the dead zone may be heated by self-gravity if sufficient material builds up. The shaded regions in the figure indicate two possible icy zones within the disc: (1) an outer icy region far from the star and (2) an inner icy region located within the dead zone, closer to the star.}
    \label{fig:sketch}
\end{figure*}

The layered disc model has different forms of viscosity  depending upon the temperature and surface density in the disc, as shown in the sketch in Fig.~\ref{fig:sketch}.  Two parameters determine the disc structure: these are the critical temperature, $T_{\rm crit}$, and the critical surface density, $\Sigma_{\rm crit}$, that we describe here.

The critical temperature is the temperature at which thermal ionization becomes important, $T_{\rm crit} = 800\,\rm K$ \citep{Gammie96, Armitage2001, Martin11}. At this temperature, the ionization fraction increases significantly due to the collisional ionization of Potassium \citep{Umebayashi1983}.  While the value for this critical temperature could be somewhat larger \citep[up to $1600\,\rm K$, e.g.][]{Zhu09}, the results we present correspond to a lower limit for the extent of the dead zone. We consider larger values and discuss this further in Section~\ref{conc}.

The critical surface density is the surface density of the surface layers that are ionized by cosmic rays or X-rays from the central star. This parameter is not well determined. If cosmic rays are the dominant source of ionization then $\Sigma_{\rm crit}\approx 200\,\rm g\, cm^{-2}$ \citep{Gammie96,Fromang2002}. However, cosmic rays may be swept away by a jet or outflow from the star \citep{Skilling1976,Cesarsky1978}, leaving X-rays from the star as the dominant ionization source. The critical surface density in this case may be significantly smaller \citep{Matsumura2003, Bai2009,Fujii2011}. Dust grains can both shield the disc from radiation and  act as a pathway for recombination \citep[e.g.][]{Woitke2016}. Given the complex nature of this parameter, we choose to keep our model simple and take $\Sigma_{\rm crit}$ to have a constant value with radius and time \citep[e.g.][]{Bourdarot2023}. In order to understand its effect on the dead zone structure and evolution, we vary the critical surface density from $2\,\rm g\,cm^{-2}$ to $200\,\rm g\,cm^{-2}$.

The disc is fully MRI-active if it satisfies one of two criteria:
\begin{enumerate}
    \item The mid-plane temperature of the disc ($T_{\rm c}$) is greater than the critical temperature ($T_{\rm crit}$), $T_{\rm c}> T_{\rm crit}$.
    \item The surface density of the disc ($\Sigma$) is less than the critical value ($\Sigma_{\rm crit}$), $\Sigma < \Sigma_{\rm crit}$. 
\end{enumerate}
The first criteria, on the mid-plane temperature, is satisfied in the inner parts of the disc.  The second criteria, on the surface density, is satisfied in the outer parts of the disc, where the surface density is low.  On the other hand, if neither criteria is satisfied, then the disc is layered.

We define two surface densities at all radii in the disc,  the surface density in the MRI-active layers, $\Sigma_{\rm m}$, and the surface density in the mid-plane layer, $\Sigma_{\rm g}$.  In the MRI-active regions, $\Sigma_{\rm m} = \Sigma$ and $\Sigma_{\rm g}= 0$. On the other hand, in the layered disc region, the active surface layers have $\Sigma_{\rm m} = \Sigma_{\rm crit}$ and the mid-plane layer has surface density $\Sigma_{\rm g} = \Sigma - \Sigma_{\rm m}$.

\subsection{Disc surface density evolution}

The material in a disc around a star with mass $M$  is in Keplerian rotation with angular frequency $\Omega = \sqrt{GM/R^3}$ at an orbital radius $R$.
The evolution of the surface density, $\Sigma(R,t)$, is determined by mass and angular momentum conservation, given by 
\begin{equation} \label{eq:1}
\frac{\partial \Sigma}{\partial t}= \frac{3}{R} \frac{\partial}{\partial R} \left(R^{1/2} \frac{\partial}{\partial R} [(\nu_{\rm m}  \Sigma_{\rm m} + \nu_{\rm g}  \Sigma_{\rm g} )R^{1/2}]\right)
\end{equation}
\citep{Pringle1981,Gammie96,Armitage2001}, where $\nu_{\rm m}$ is the viscosity in the MRI-active layers and
$\nu_{\rm g}$ is the viscosity due to turbulence associated with self-gravity of the gas that may act outside the MRI layers.

For the viscosity in the MRI-active layers, we adopt an alpha prescription given by
\begin{equation} \label{eq:nu_m}
\nu_{\rm m} = \alpha_{\rm m} \frac{{c_{\rm m}}^2}{\Omega},
\end{equation}
where $c_{\rm m}=\sqrt{\mathcal{R} T_{\rm m} / \mu}$ is the sound speed in the MRI-active layer with temperature $T_{\rm m}$ and $\alpha_{\rm m}$ is a dimensionless parameter measuring the efficiency of angular momentum transport. 

Within the dead zone, the disc becomes self-gravitating when the \cite{Toomre} parameter defined as
\begin{equation} \label{eq:Q}
Q = \frac{c_{\rm g} \Omega}{\pi G \Sigma},
\end{equation}
becomes less than its critical value $Q_{\rm crit }=2$. The sound speed at the disc mid-plane is $c_{\rm g}=\sqrt{\mathcal{R}T_{\rm c} / \mu}$. 
If $Q<Q_{\rm crit}$ the viscosity due to turbulence associated with self-gravity is
\begin{equation} \label{eq:nu_g}
\nu_{\rm g} = \alpha_{\rm g} \frac{{c_{\rm g}}^2}{\Omega},
\end{equation}
where
\begin{equation} \label{eq:alpha_g}
 \alpha_{\rm g} = 0.01\left( \frac{{Q_{\rm crit}}^2}{Q^2} - 1 \right),
\end{equation}
 and $\nu_{\rm g}$ = 0 otherwise \citep{Lin1,Lin1990}. \\

\subsection{Disc temperature evolution}

The temperature at the disc mid-plane  evolves according to the simplified energy equation 
\begin{equation} \label{eq:Tc}
\frac{\partial T_{\rm c}}{\partial t}= \frac{2(Q_+ - Q_-)}{c_{\rm p} \Sigma} 
\end{equation}
\citep{Pringle1986,Canizzo1993}, where $c_{\rm p}$ is the disc specific heat, which for temperatures $T_{\rm c} \sim 10^3 K$ is given by $c_{\rm p} = 2.7 \mathcal{R}/ \mu$ \citep{Armitage2001} where $\mathcal{R}$ is the gas constant and $\mu = 2.3$ is the gas mean molecular weight. The local heating owing to viscous dissipation is given by
\begin{equation} \label{eq:Qplus}
Q_+ = \frac{9}{8}   \Omega^2 (\nu_{\rm m}  \Sigma_{\rm m} + \nu_{\rm g}  \Sigma_{\rm g} ).
\end{equation}
For the cooling rate, we assume that each annulus of the disc radiates as a black body at temperature $T_{\rm e}$, so that
\begin{equation} \label{eq:Qminus}
Q_- = \sigma {T_{\rm e}}^4,
\end{equation}
where $\sigma$ is the Stefan-Boltzmann constant and $T_{\rm e}$ is the surface temperature. Thus, equation~(\ref{eq:Tc}) in a steady state has the solution
\begin{equation}\label{eq:ss}
\sigma {T_{\rm e}}^4 =  \frac{9}{8}  \Omega^2 (\nu_{\rm m}  \Sigma_{\rm m} + \nu_{\rm g}  \Sigma_{\rm g} ).
\end{equation}

The mid-plane disc temperature for an optically thick disc in thermal equilibrium is obtained by considering the energy balance in a layered model above the disc mid-plane. The MRI surface layer contains the surface density $\Sigma_{\rm m}/2$. The mid-plane layer contains the surface density $\Sigma_{\rm g}/2$. The results are that 
\begin{equation} \label{eq:Tc2}
    \sigma {T_{\rm c}}^4 =  \frac{9}{8}   \Omega^2 \left(\nu_{\rm m}  \Sigma_{\rm m} \tau_{\rm m} + \nu_{\rm g}  \Sigma_{\rm g} \tau \right),
\end{equation}
and 
\begin{equation} \label{eq:Tm}
    {T_{\rm m}}^4 =  \tau_{\rm m} {T_{\rm e}}^4.
\end{equation}
The optical depth to the magnetic region is
\begin{equation} \label{eq:taum}
 \tau_{\rm m} =  \frac{3}{8} \kappa(T_{\rm m}) \Sigma_{\rm m}/2,
\end{equation}
and the optical depth within the mid-plane region is
\begin{equation} \label{eq: taug}
 \tau_{\rm g} =  \frac{3}{8} \kappa(T_{\rm c}) \Sigma_{\rm g}/2,
\end{equation}
where
\begin{equation} \label{eq:tau}
 \tau =  \tau_{\rm m} + \tau_{\rm g}.
\end{equation}
The behavior of the disc depends primarily on the opacity near the transition temperature $T_{\rm crit}$, for which the fit is
\begin{equation} \label{eq:kappa}
\kappa = 0.02T^{0.8}\, \rm cm^2\, g^{-1}
\end{equation}
\citep{Armitage2001}, where $T$ refers to the local temperature: $T_{\rm m}$ in the MRI-active layers and $T_{\rm c}$ in the mid-plane region. From equations~(\ref{eq:Qminus}),~(\ref{eq:ss}),~(\ref{eq:Tc2}) and~(\ref{eq:tau}), we derive an expression for the cooling function which is applied even when the disc is not in thermal equilibrium
\begin{equation}\label{eq:Qminus2}
Q_{-} = \frac{1}{\tau} \left( \sigma T_{\rm c}^{4} + \frac{9}{8} \Omega^2 \nu_{\rm m} \Sigma_{\rm m} \tau_{\rm g} \right)
\end{equation}   
\citep{Martin11}. 

\subsection{Irradiation from the star}
\label{sec:Irr}

In the innermost parts of the disc, the irradiation from the star  may dominate the temperature of the disc over the viscous heating. The flux of radiation is
\begin{equation}
 F_{\rm irr} =  \sigma T_*^{4} \frac{\alpha_{\rm irr}}{2} \left( \frac{R_*}{R} \right)^{2},
\end{equation}
where $T_*$ is the effective temperature of the star, $R_*$ is the radius of the star and
\begin{equation}
 \alpha_{\rm irr} = 0.005 \left( \frac{R}{AU} \right)^{-1} + 0.05 \left( \frac{R}{AU} \right)^{2/7}
\end{equation}
\citep{Kenyon87, Chiang97}. The irradiation temperature is given by
\begin{equation}
 T_{\rm irr} = \left( \frac{F_{\rm irr}}{\sigma} \right)^{1/4}
 \label{tirr}
\end{equation}
\citep{Chiang97}. If the mid-plane temperature of the disc drops below the irradiation temperature, then we assume that the disc is isothermal and set $ T_{\rm c} = T_{\rm m}=T_{\rm e}=T_{\rm irr}$. 
The stellar temperatures and radii for stellar mass are shown in Table~\ref{tab:params} \citep[see also e.g.][]{Hasegawa10}.

 \begin{figure*}
    \centering
    \begin{subfigure}[b]{0.47\textwidth}
        \includegraphics[width=\textwidth]{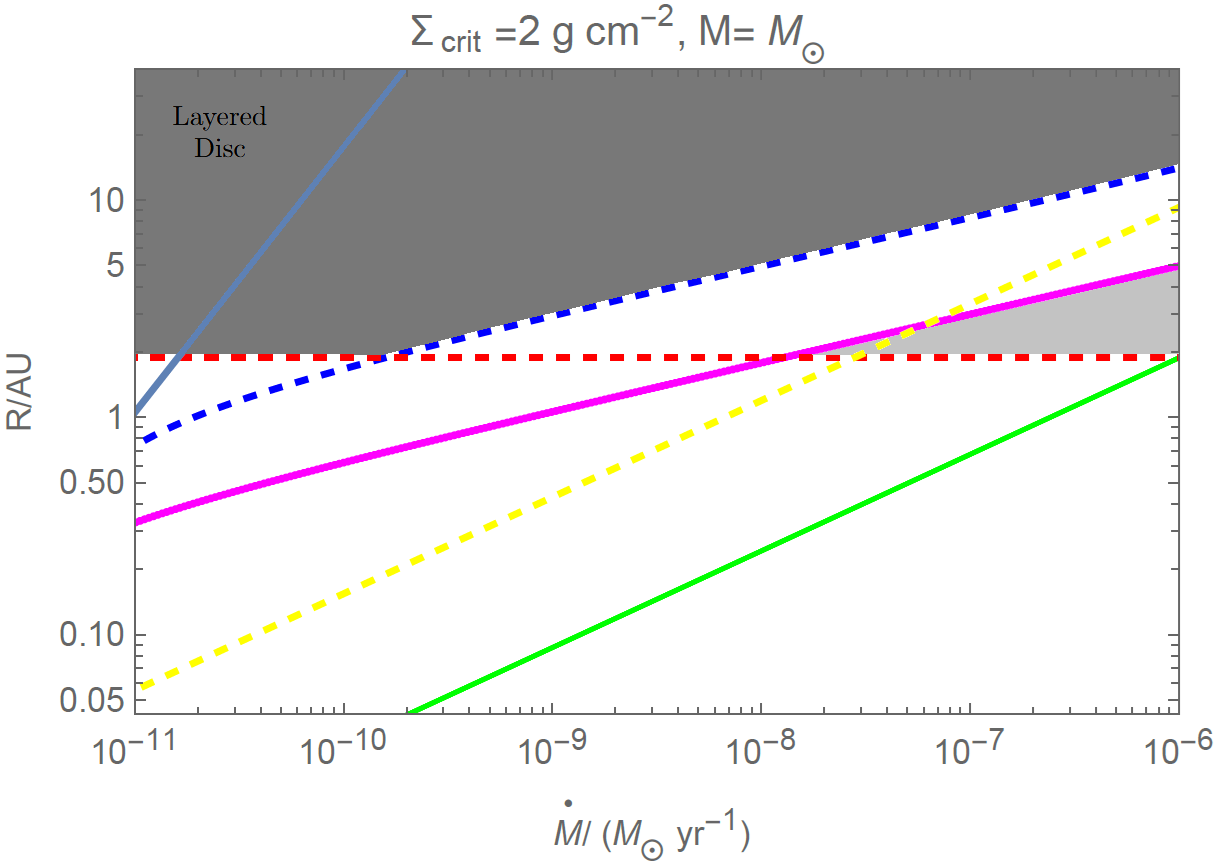}
    \end{subfigure}
    \hfill
    \begin{subfigure}[b]{0.47\textwidth}
        \includegraphics[width=\textwidth]{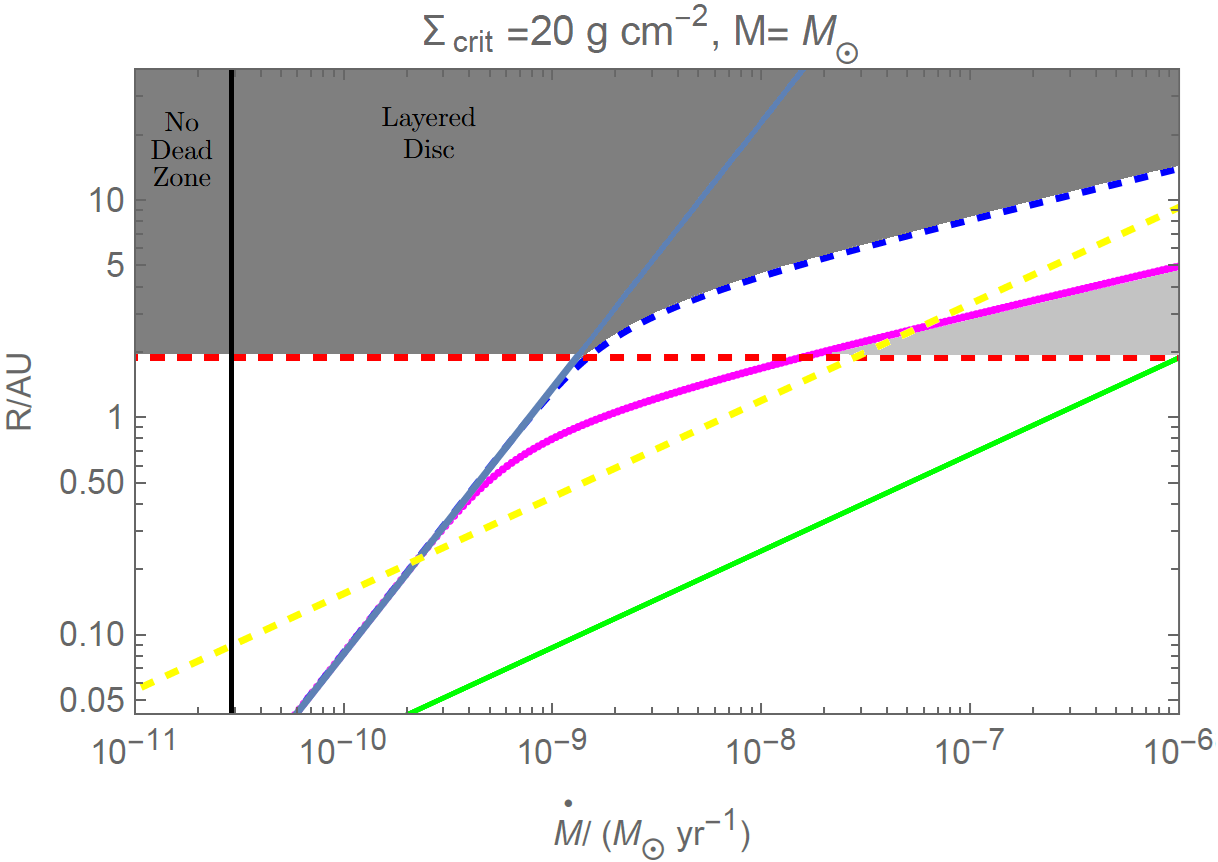}
    \end{subfigure}
    \hfill
    \begin{subfigure}[b]{0.47\textwidth}
        \includegraphics[width=\textwidth]{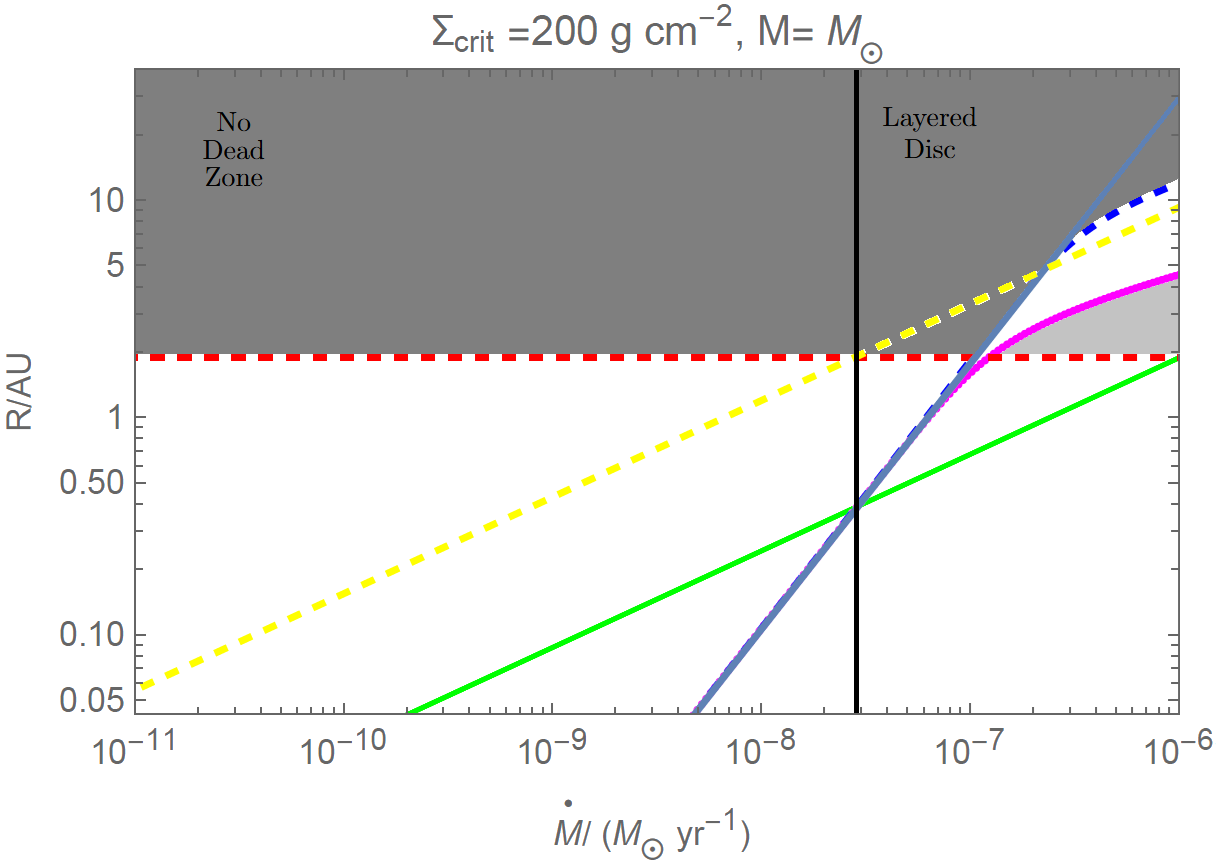}
    \end{subfigure}
    \hfill
    \makebox[200pt][l]{
    \begin{minipage}[b]{0.37\textwidth}
        \raisebox{15pt}{
            \includegraphics[width=\textwidth]{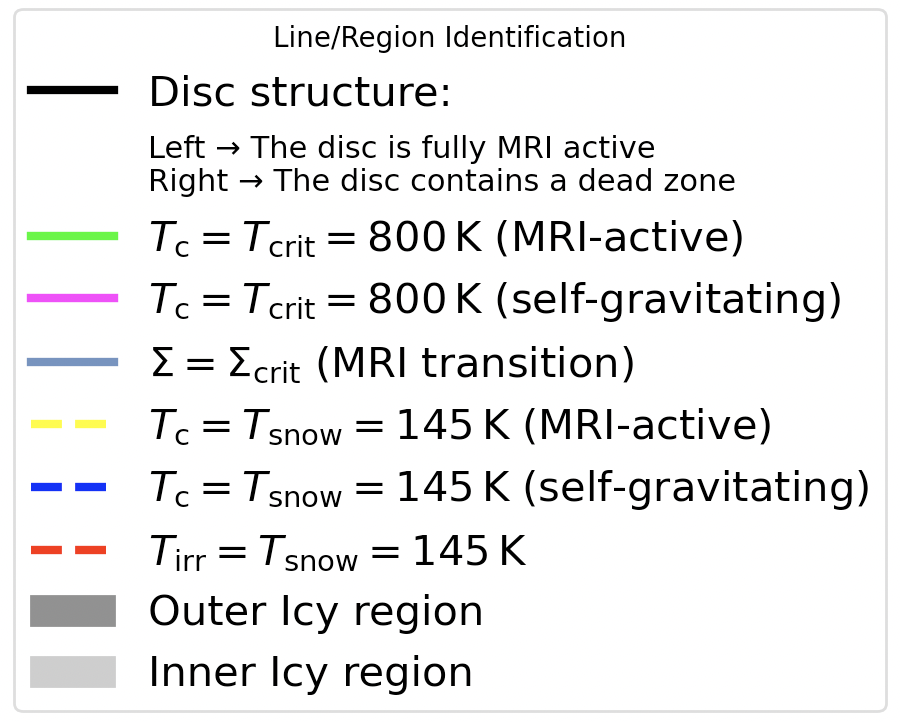}
        }
    \end{minipage}}

         \caption{The structure of the disc as a function of the steady-state accretion rate through the disc for stellar mass $M=1\,\rm M_\odot$. The shaded regions show where the disc can be icy. The critical surface density is $\Sigma_{\rm crit}=2\,\rm g\, cm^{-2}$ (upper left), $\Sigma_{\rm crit}=20\,\rm g\, cm^{-2}$ (upper right) and $\Sigma_{\rm crit}=200\,\rm g\, cm^{-2}$ (lower). Solid lines indicate boundaries of the disc structure. The vertical black line shows where the disc transitions from a fully turbulent disc without a dead zone (low accretion rates) to a layered disc (high accretion rates). For the lowest $\Sigma_{\rm crit}$, the disc is always layered. The solid green lines show the radius at which $T_{\rm c}=T_{\rm crit}=800\,\rm K$ for a fully MRI-active disc (see Section \ref{sec:FMRI}). The solid light blue lines show the critical radius for the transition to the MRI branch in the outer part of the disc (where $\Sigma=\Sigma_{\rm crit}$, see Section \ref{sec:FMRI}). The magenta lines show where $T_{\rm c}=T_{\rm crit}=800\, \rm K$ in the self-gravitating disc (see Section~\ref{sec:SGSS}.). The dashed lines indicate snow line radii. The dashed red lines show where the $T_{\rm irr}=T_{\rm snow}=145\,\rm K$ (see Section \ref{subsec:irr dics}). The dashed yellow lines show the radius at which $T_{\rm c}=T_{\rm snow}=145$K for a fully MRI-active disc.  The dashed blue lines show the snow line radius in a self-gravitating disc.   
         The outer dark-shaded icy region is bounded by steady disc solutions and does not evolve in time. The light-shaded region can be icy, but since the disc is not in a steady state in a dead zone region, the icy region may evolve in time (see Section \ref{sec:icy}).} 
    \label{fig:vallet1}
\end{figure*}

 \begin{figure*}
    \centering
    \begin{subfigure}[b]{0.47\textwidth}
        \includegraphics[width=\textwidth]{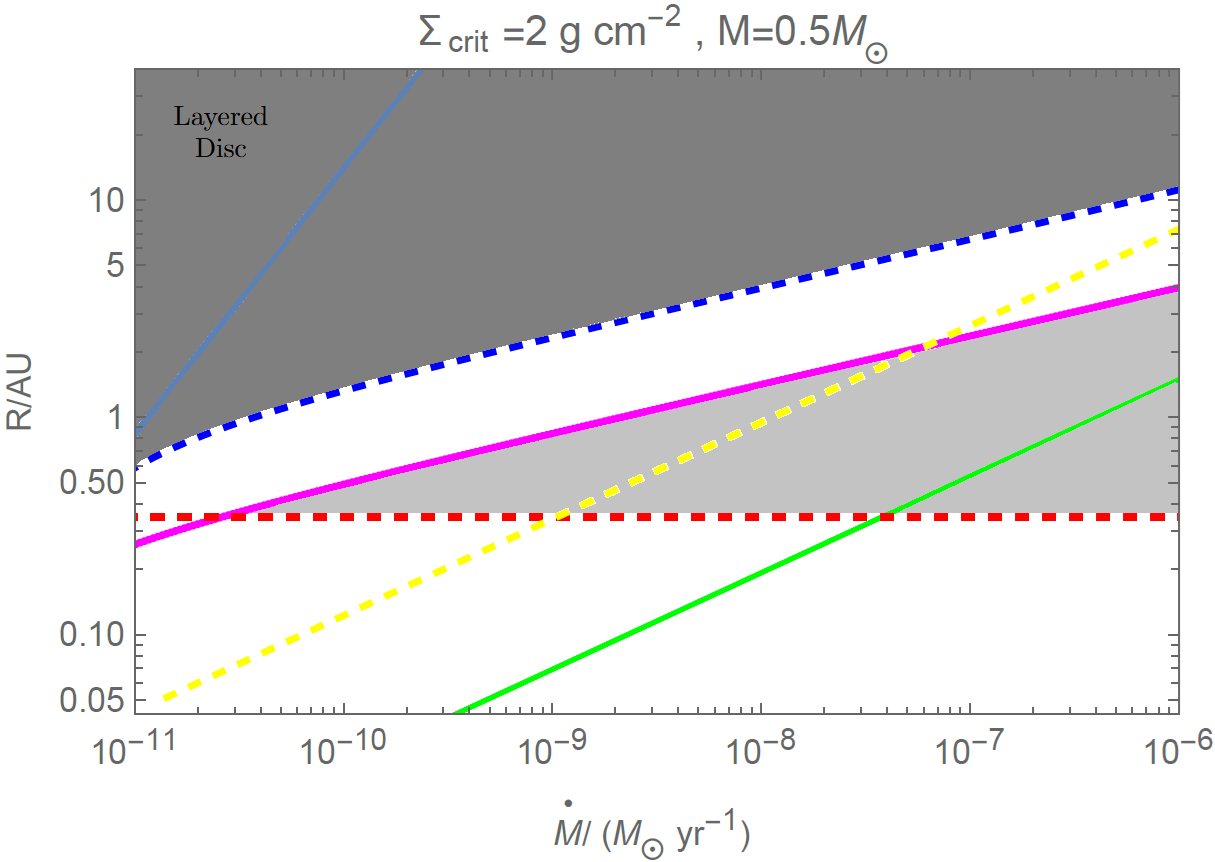}
    \end{subfigure}
    \hfill
    \begin{subfigure}[b]{0.47\textwidth}
        \includegraphics[width=\textwidth]{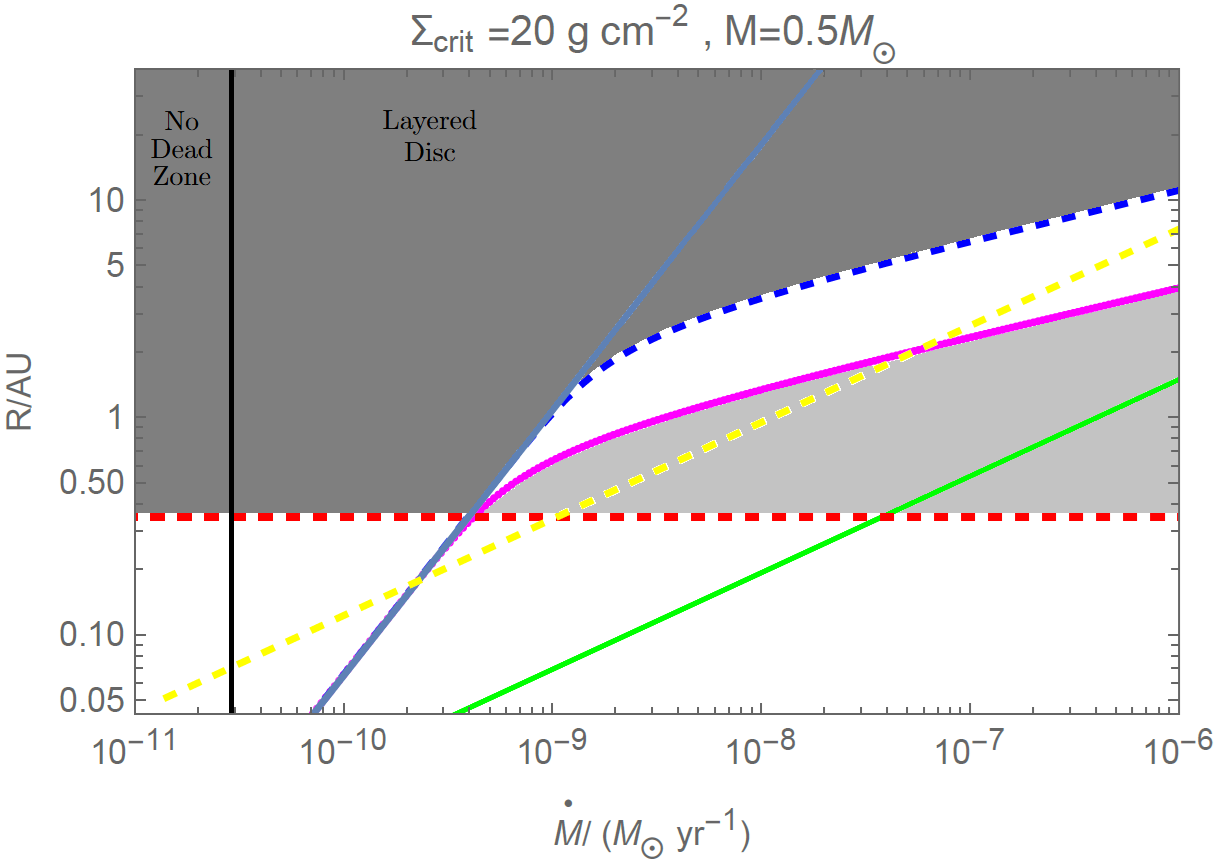}
    \end{subfigure}
    \hfill
    \begin{subfigure}[b]{0.47\textwidth}
        \includegraphics[width=\textwidth]{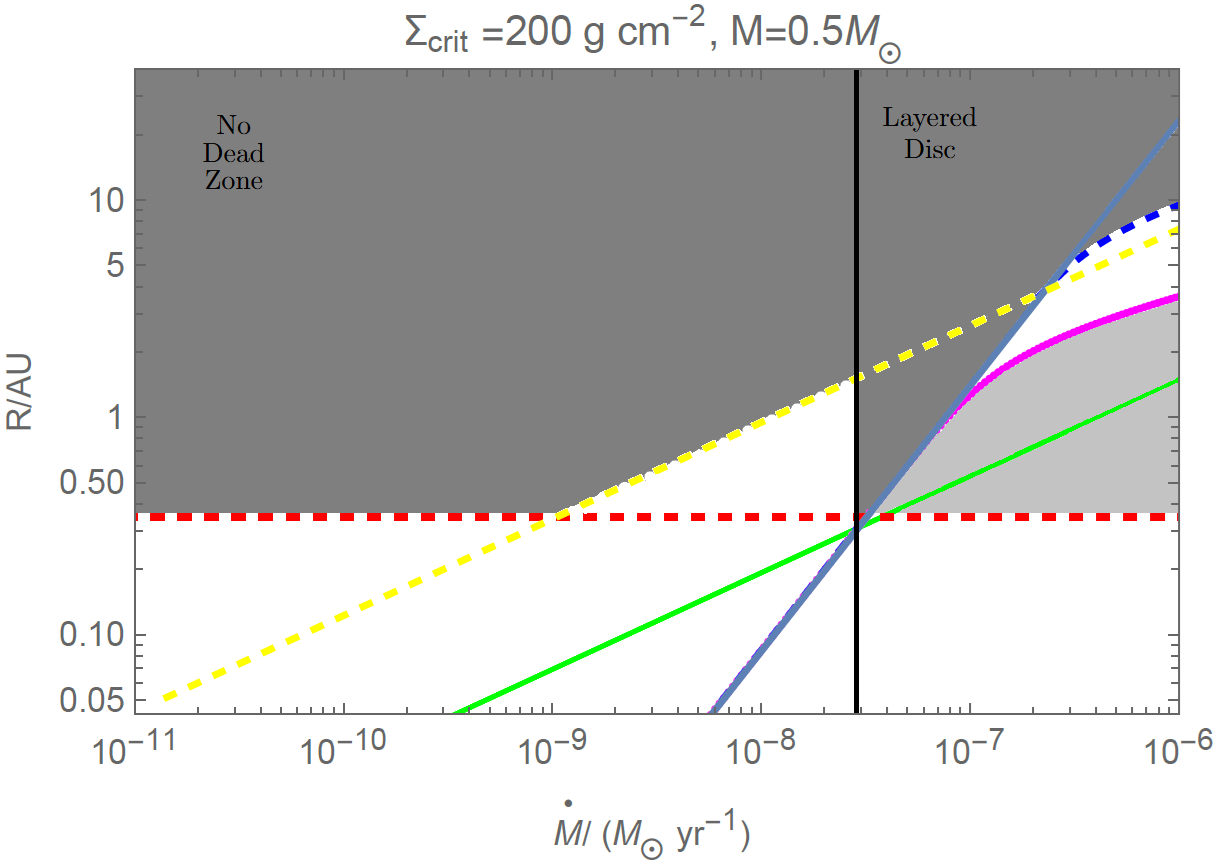}
    \end{subfigure}
    \hfill
    \makebox[200pt][l]{
    \begin{minipage}[b]{0.37\textwidth}
        \raisebox{15pt}{
            \includegraphics[width=\textwidth]{images/label.png}
        }
    \end{minipage}}

    \caption{Same as Fig.~\ref{fig:vallet1} except $M=0.5\,\rm M_\odot$.}
    \label{fig:vallet05}
\end{figure*}

 \begin{figure*}
    \centering
    \begin{subfigure}[b]{0.47\textwidth}
        \includegraphics[width=\textwidth]{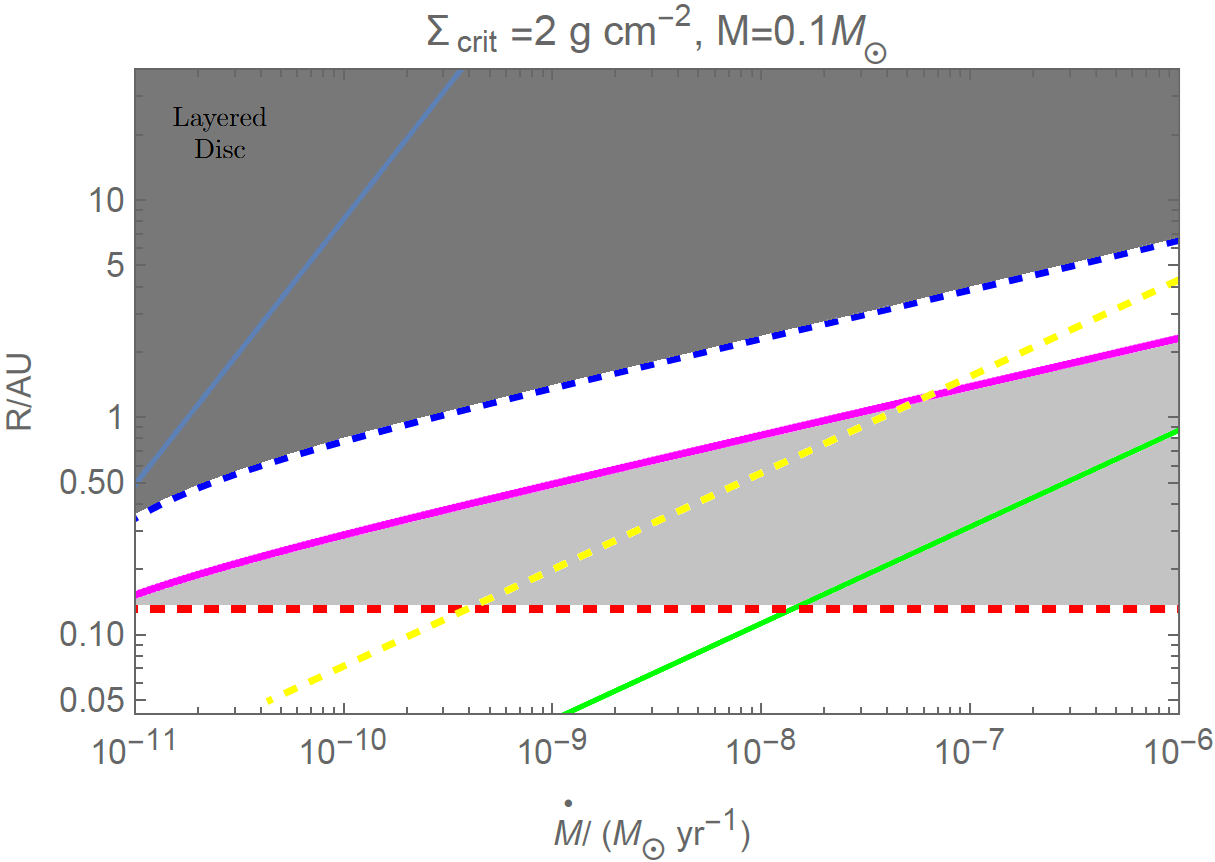}
    \end{subfigure}
    \hfill
    \begin{subfigure}[b]{0.47\textwidth}
        \includegraphics[width=\textwidth]{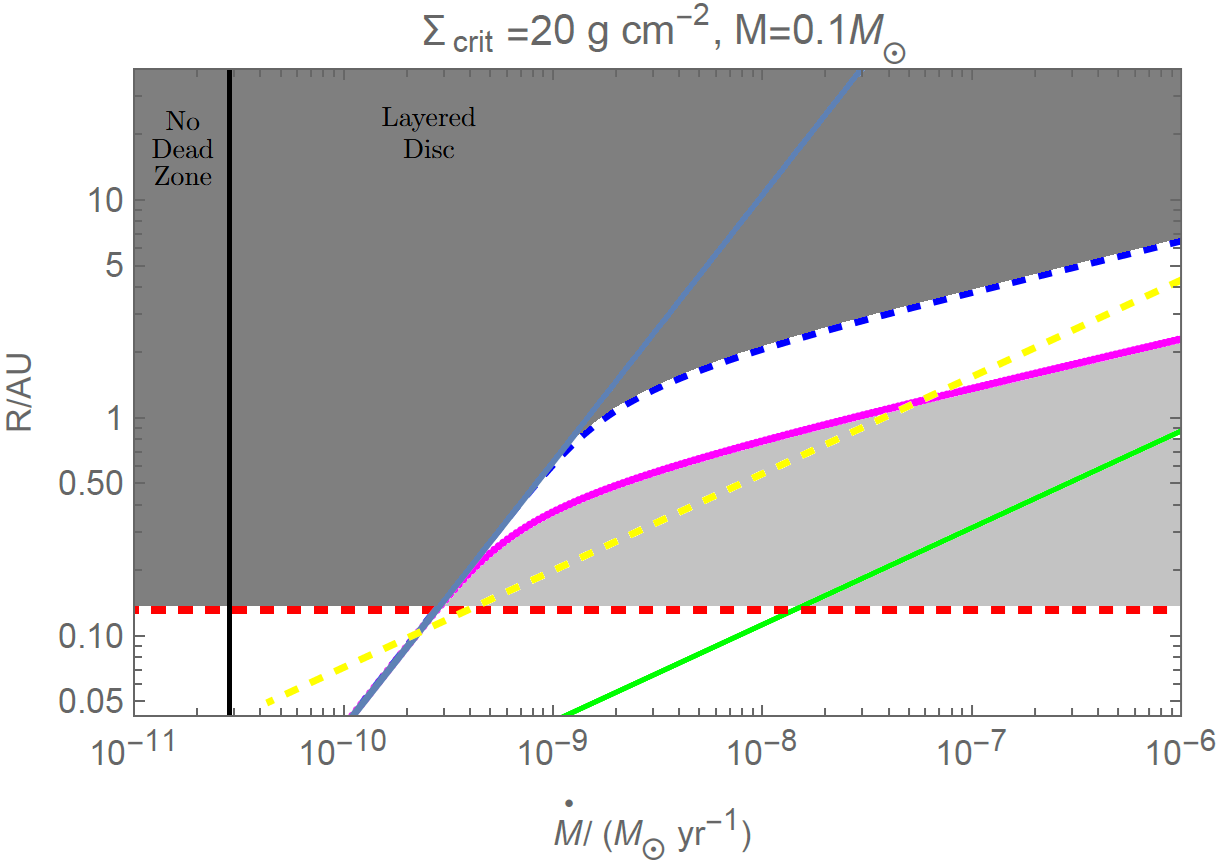}
    \end{subfigure}
    \hfill
    \begin{subfigure}[b]{0.47\textwidth}
        \includegraphics[width=\textwidth]{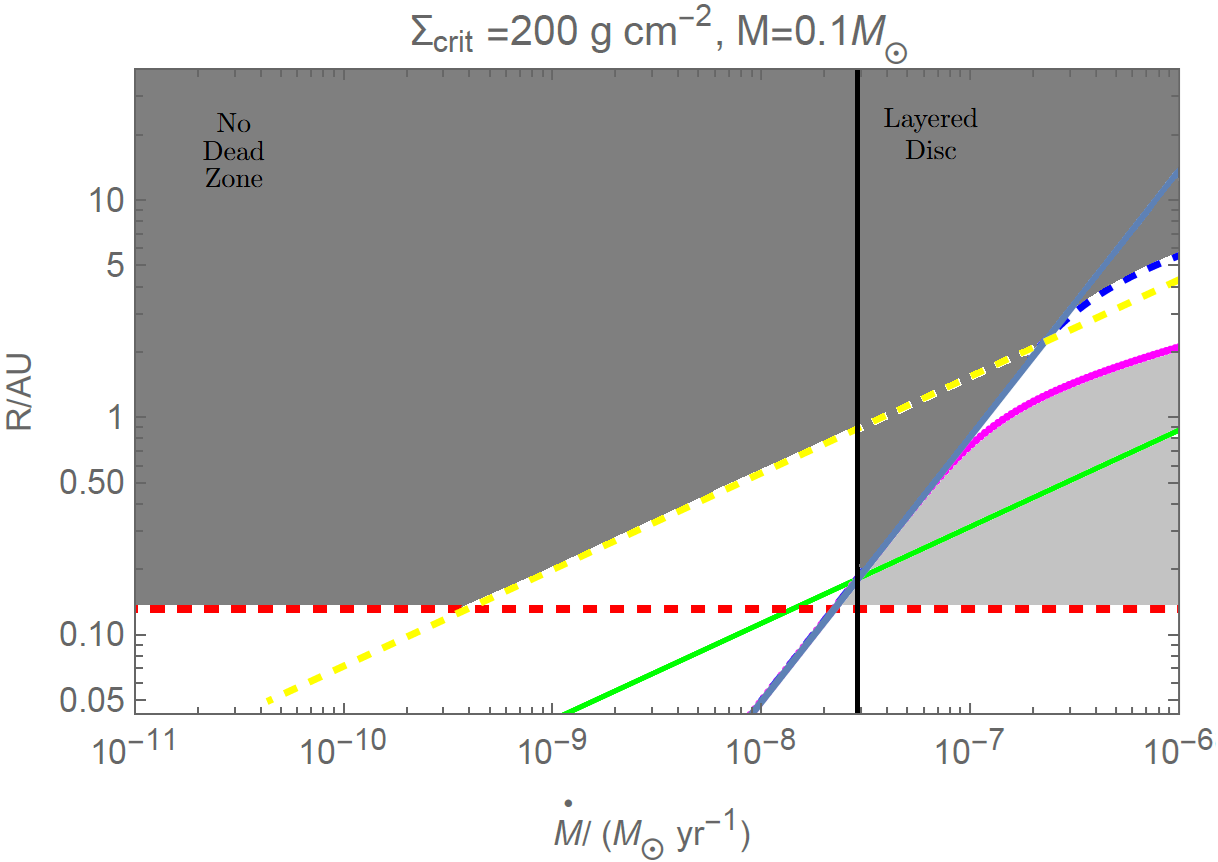}
    \end{subfigure}
    \hfill
    \makebox[200pt][l]{
    \begin{minipage}[b]{0.36\textwidth}
        \raisebox{15pt}{
            \includegraphics[width=\textwidth]{images/label.png}
        }
    \end{minipage}}

    \caption{Same as Fig.~\ref{fig:vallet1} except $M=0.1\,\rm M_\odot$.}
    \label{fig:vallet01}
\end{figure*}

\section{Steady-state disc solutions}
\label{sec:result}

In this section, we consider steady-state disc solutions. These enable us to place bounds on the possible icy regions in the disc.  We consider solutions for an irradiation-dominated disc, a fully turbulent disc, and a disc with a self-gravitating dead zone. A steady-state disc can have these different solutions in different radial regions. We determine the location of the snow line radius for each case. The snow line radius is defined by a temperature, $T_{\rm snow}$, which falls within the range of  $145\,\rm K$ to $170\,\rm K$ \citep{Lecar1, Hayashi} depending on the partial pressure of nebular water vapour. To adopt a more conservative estimate for the extent of icy regions, we take $T_{\rm snow}=145\,\rm K$ in our model. 

In a steady state, the mass accretion rate through the disc is constant  at all radii. Far from radius of the central body, equation (\ref{eq:1}) yields the following solution
\begin{equation} \label{eq:mss}
\dot{M} = 3 \pi (\nu_{\rm m} \Sigma_{\rm m} + \nu_{\rm g} \Sigma_{\rm g}).
\end{equation}
Thus, from equations (\ref{eq:ss}) and (\ref{eq:mss}) we determine the steady-state surface temperature of the disc
\begin{equation} \label{eq:Tess}
T_{\rm e} = \left( \frac{3 \dot{M} \Omega^2}{8 \pi \sigma} \right)^{\frac{1}{4}}.
\end{equation}
Figs.~\ref{fig:vallet1} ($M=1.0\,\rm M_\odot$), \ref{fig:vallet05} ($M=0.5\,\rm M_\odot$), and \ref{fig:vallet01} ($M=0.1\,\rm M_\odot$) present the full set of solutions as a function of the steady-state accretion rate through the disc.
The shaded areas indicate the possible icy regions of a disc, where the mid-plane temperature $T_c$ can drop below the snow line temperature. A detailed explanation of these figures is provided in the following sub-sections. Note that these results are consistent with those of \cite{Vallet23} in the limit $\Sigma_{\rm crit}\rightarrow 0$.

\subsection{Irradiation-dominated disc}
\label{subsec:irr dics}
The snow line radius in an irradiation-dominated disc is determined by setting $T_{\rm irr}=T_{\rm snow}=145 \, \rm K$ using equation~(\ref{tirr}). The red dashed lines in Figs.~\ref{fig:vallet1} ($M=1.0\,\rm M_\odot$), \ref{fig:vallet05} ($M=0.5\,\rm M_\odot$), and \ref{fig:vallet01} ($M=0.1\,\rm M_\odot$) indicate this snow line radius. For a solar-mass star, the snow line is located at $R_{\rm snow} = 1.88\,\rm au$. For an M-dwarf with mass $M=0.5\,\rm M_{\odot}$, it lies at $R_{\rm snow} = 0.35\,\rm au$, and for $M=0.1\,\rm M_{\odot}$, at $R_{\rm snow} = 0.10\,\rm au$.
 These values are independent of the disc accretion rate or the critical surface density.

\subsection{Fully turbulent disc}
\label{sec:FMRI}

The structure of a fully turbulent disc is determined by solving equations (\ref{eq:Tc2})  and (\ref{eq:mss}) with $\Sigma_{\rm m}=\Sigma$ and $\Sigma_{\rm g}=0$. 
In Figs.~\ref{fig:vallet1} ($M=1.0\,\rm M_\odot$), \ref{fig:vallet05} ($M=0.5\,\rm M_\odot$), and \ref{fig:vallet01} ($M=0.1\,\rm M_\odot$), the green lines indicate the radius where $T_{\rm c} = T_{\rm crit}$. These lines are independent of the critical surface density value.  Below the  green lines, there exists a fully turbulent steady-state solution with $T_{\rm c}> T_{\rm crit}$. However, at larger radii, there may be no solution that maintains a sufficiently high temperature to prevent the formation of a dead zone. 

In a disc with a dead zone, the outer parts may be fully MRI-active if $\Sigma<\Sigma_{\rm crit}$. The outer radius at which the disc is fully MRI-active is found by solving equations~(\ref{eq:Tc2})  and~(\ref{eq:mss}) with $\Sigma_{\rm m}=\Sigma_{\rm crit}$ and  $\Sigma_{\rm g} = 0$. The light blue solid lines indicate the outer radius where the transition to the MRI-active branch occurs. The dead zone, therefore, can form in discs where the blue line is above the green line. At the accretion rate where these lines cross, we show a vertical black line. This denotes the transition from a disc with a dead zone to a fully MRI-active disc.  To the right of the black line, there is no fully MRI-active steady state solution. To the left of this black line, there exists a fully MRI-active solution at all radii in the disc. For larger $\Sigma_{\rm crit}$, the transition where the disc becomes fully MRI-active occurs at a larger accretion rate. 

The yellow dashed lines show where $T_{\rm c} = T_{\rm snow}$ in the fully MRI-active solution. However, note that this line is only relevant in discs that are fully MRI-active everywhere (i.e., to the left of the vertical solid black line). The snow line radius is closer to the star at lower accretion rates. 

\subsection{Self-gravitating disc}
\label{sec:SGSS}
 Forming a dead zone in the disc requires the critical surface density $\Sigma_{\rm crit}$ to be sufficiently low, so there is no steady-state solution throughout the disc. The material builds up in the dead zone until the outer parts become self-gravitating. We find steady self-gravitating disc solutions by solving equations (\ref{eq:Tc2})  and (\ref{eq:mss}).
 The blue dashed lines in Figs.~\ref{fig:vallet1} ($M=1.0\,\rm M_\odot$), \ref{fig:vallet05} ($M=0.5\,\rm M_\odot$), and \ref{fig:vallet01} ($M=0.1\,\rm M_\odot$) show the snow line radius in a disc with a self-gravitating dead zone.  The snow line in a self-gravitating disc can extend much farther out compared to a fully turbulent disc \citep[see also][]{Martin2012,Martin2013}. The magenta lines show where the steady solution has $T_{\rm c}=T_{\rm crit}$.  This is the minimum radius for this steady solution, as reaching the critical temperature at this point would trigger the MRI. 
 
\subsection{Icy Regions}
\label{sec:icy}

In Figs.~\ref{fig:vallet1} ($M=1.0\,\rm M_\odot$), \ref{fig:vallet05} ($M=0.5\,\rm M_\odot$), and \ref{fig:vallet01} ($M=0.1\,\rm M_\odot$), the shaded areas indicate the possible icy regions of a disc containing a dead zone, where the mid-plane temperature $T_c$ can drop below the snow line for various steady-state accretion rates, depending on the critical surface density. The outer parts of the disc far from the star are icy, and within the dead zone, there may be an additional inner icy region. 

The outer icy region is bounded by the snow line of a self-gravitating disc (dashed blue line), the fully MRI turbulent disc (yellow dashed line), or the irradiation temperature (red dashed line). Since these are steady-state discs in the outer disc regions, the outer icy region does not change in time for a fixed accretion rate. 
Higher critical surface density leads to a fully turbulent disc for a larger range of accretion rates. Although the outer icy region may extend close to the star for accretion rates for which there is a fully turbulent disc, in-situ planet formation may be difficult without a dead zone. 

The inner icy region occurs for accretion rates where there is a dead zone in the disc. It is limited at the inner radius by the irradiation-dominated disc snow line (red dashed line) and at the outer radius by the point where the steady self-gravitating solution has a temperature $ T_{ \rm crit}$ (the magenta line). The disc temperature profile has a sharp increase in a small radial range where self-gravity begins \citep[e.g.][]{ML2013}. 

For larger $\Sigma_{\rm crit}$, the range of accretion rates for which the inner icy region exists is smaller.
This suggests that for larger values of $\Sigma_{\rm crit}$, the disc is less likely to maintain temperatures below the snow line in this region. The size of the inner icy region increases as the stellar mass decreases. Provided that $\Sigma_{\rm crit}\lesssim 20\,\rm g\, cm^{-2}$, the two M-dwarf mass stars can have extensive inner icy regions for a wide range of infall accretion rates. However, for a solar mass star, the inner icy region is only present for relatively large infall accretion rates. We note the icy regions presented here provide some bounds to the region, but we need time-dependent models in the next section to constrain this region further.

\section{Time-dependent disc evolution}
\label{sec:timedepen}

\begin{figure*}
  \centering
  \begin{subfigure}[b]{0.48\textwidth}
    \includegraphics[width=\textwidth]{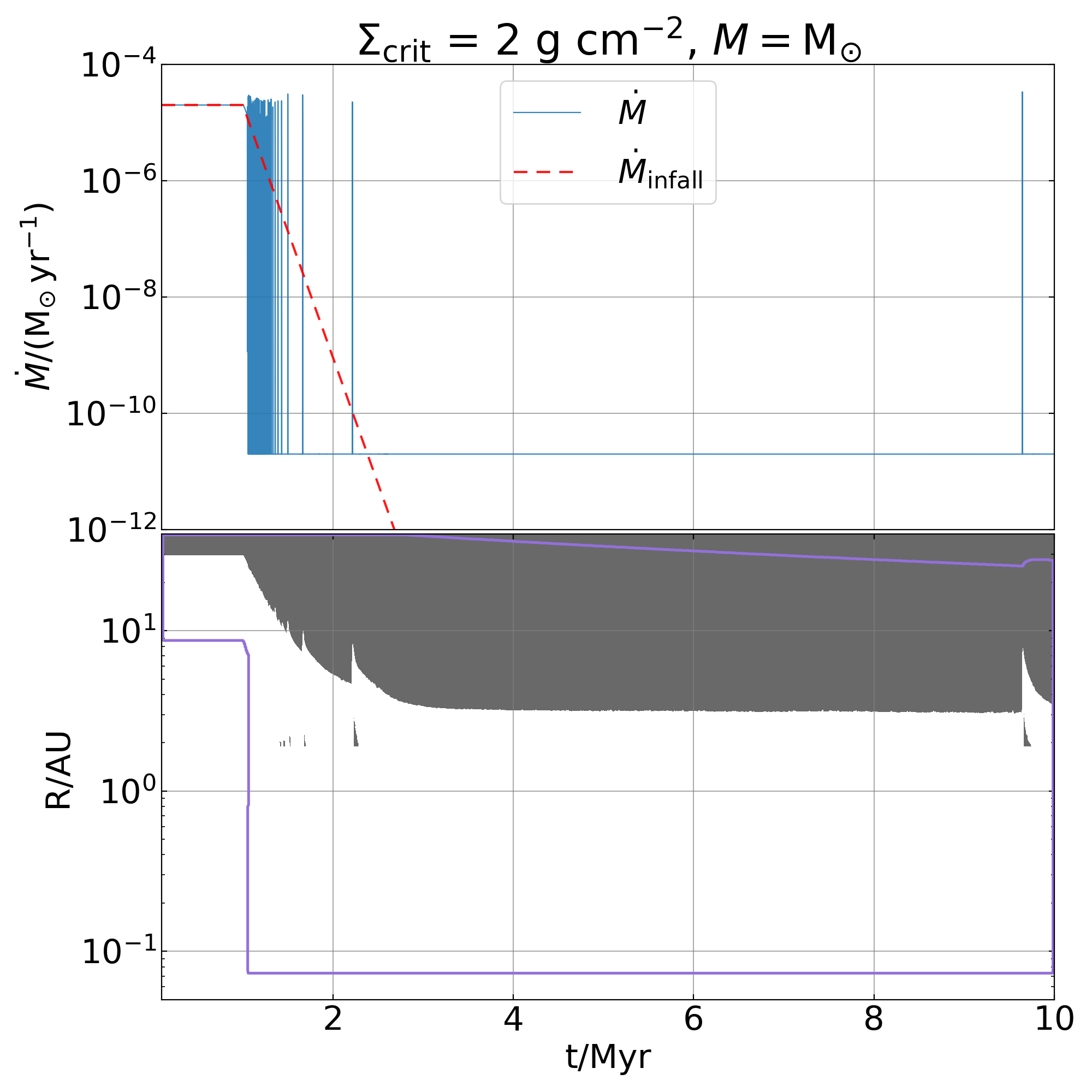}
  \end{subfigure}\hfill
  \begin{subfigure}[b]{0.48\textwidth}
    \includegraphics[width=\textwidth]{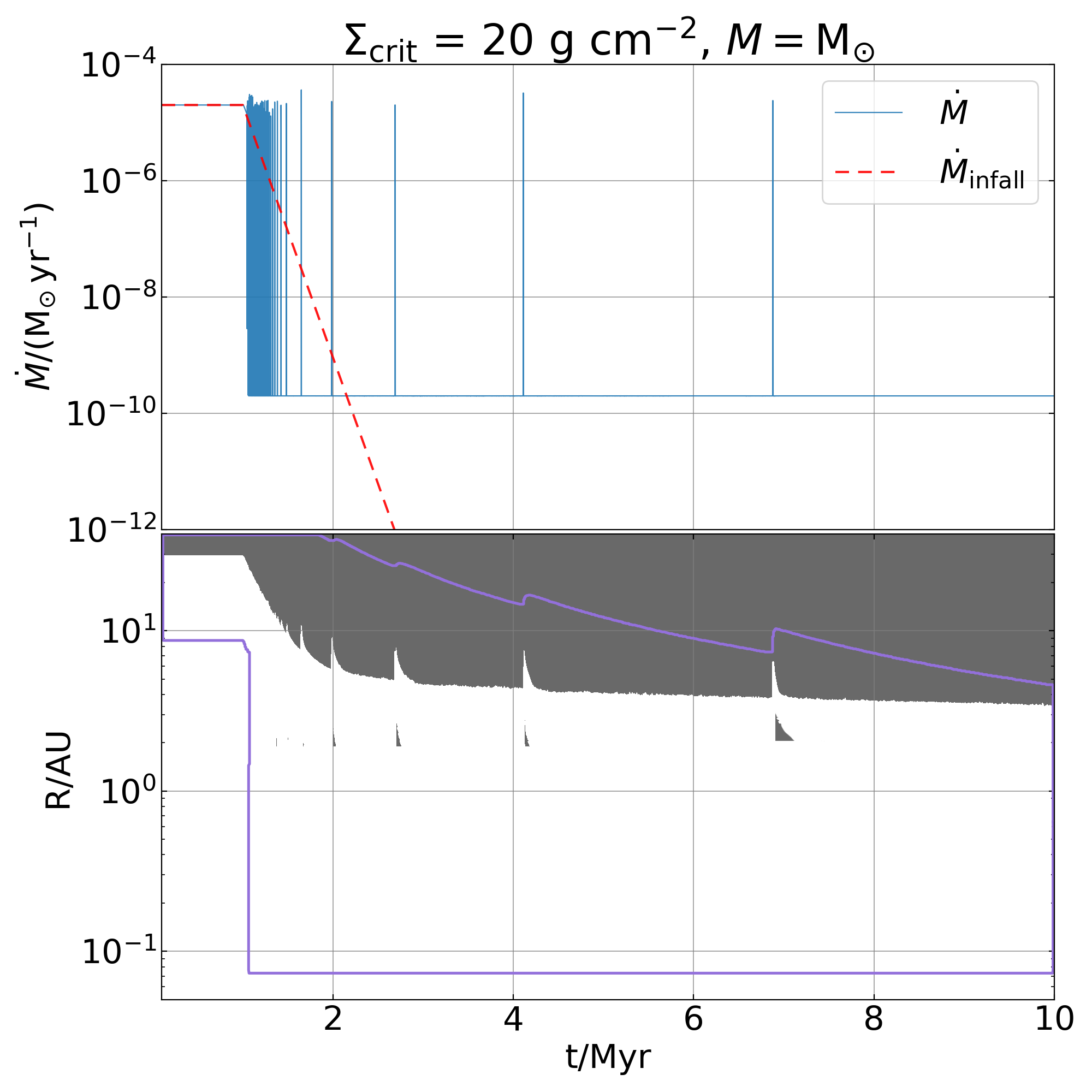}
  \end{subfigure}

  \vspace{0.6em}
  \par 
  \begin{subfigure}[b]{0.48\textwidth}
    \centering
    \includegraphics[width=\textwidth]{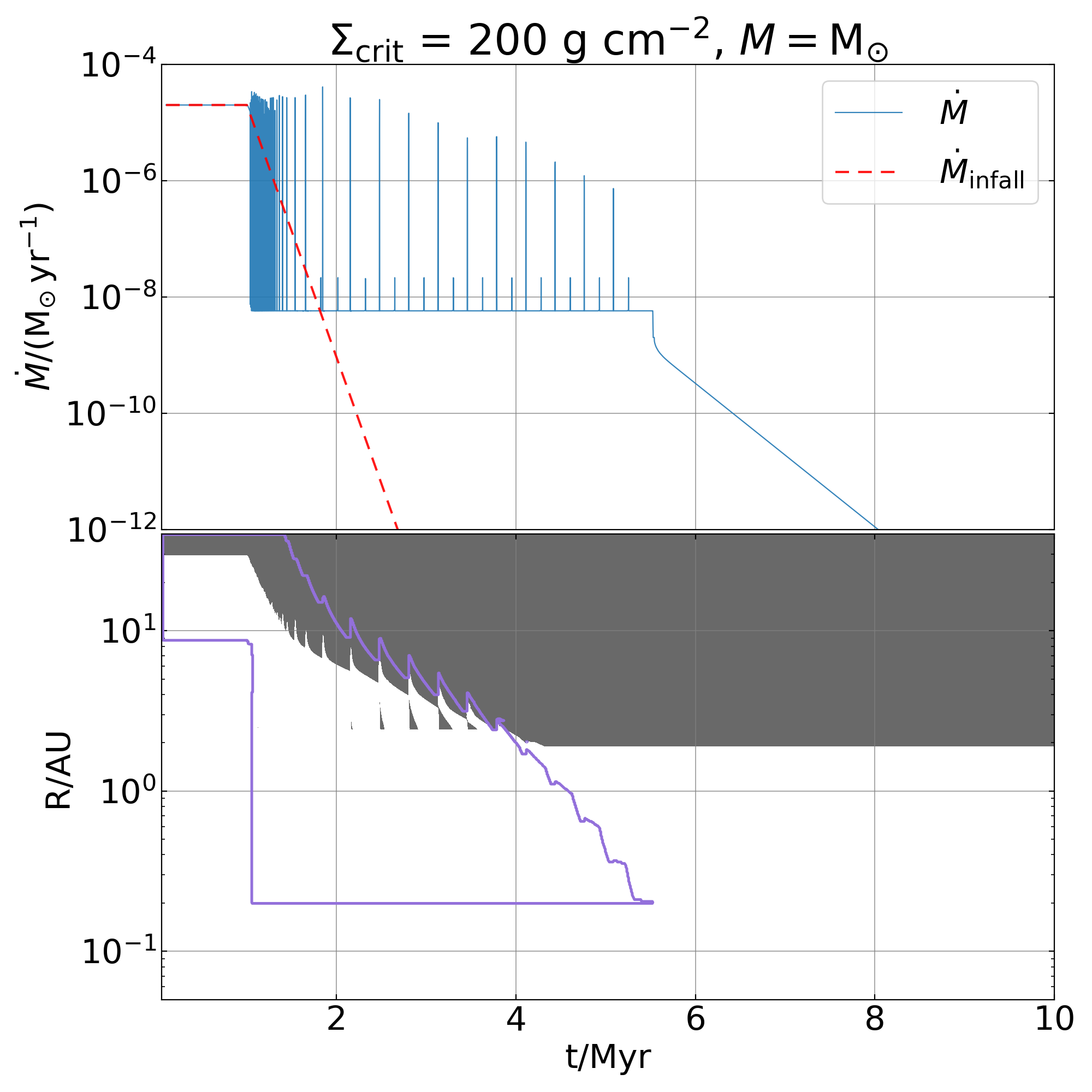}
  \end{subfigure}
    \caption{Time-dependent simulations for a disc around a star with mass $M=1\,\rm M_\odot$ with initial infall rate $\dot M_{\rm i}=2\times 10^{-5}\,\rm M_\odot \, yr^{-1}$, and  $\Sigma_{\rm crit}=2$ (top left panel), 20 (top right panel), and $200\,\rm g\,cm^{-2}$ (bottom panel).
    Top row of each panel: The accretion rate onto the star over time. The red dashed lines show the infall accretion rate given by equation~(\ref{eq:mdot}). The light blue lines show the accretion rate onto the star.     
    Bottom row of each panel: The evolution of the icy regions within the disc over time. The gray areas indicate the regions of the disc where the temperature at the mid-plane disc is low enough for ice to exist ($T_{\rm c}<T_{\rm snow}$).The region enclosed by the purple line illustrates how the extent of the dead zone evolves over time.At times where there are no purple lines, there is no dead zone, and the disc is fully turbulent.}
   \label{fig:graph1}
\end{figure*}

\begin{figure*}
  \centering
  \begin{subfigure}[b]{0.48\textwidth}
    \includegraphics[width=\textwidth]{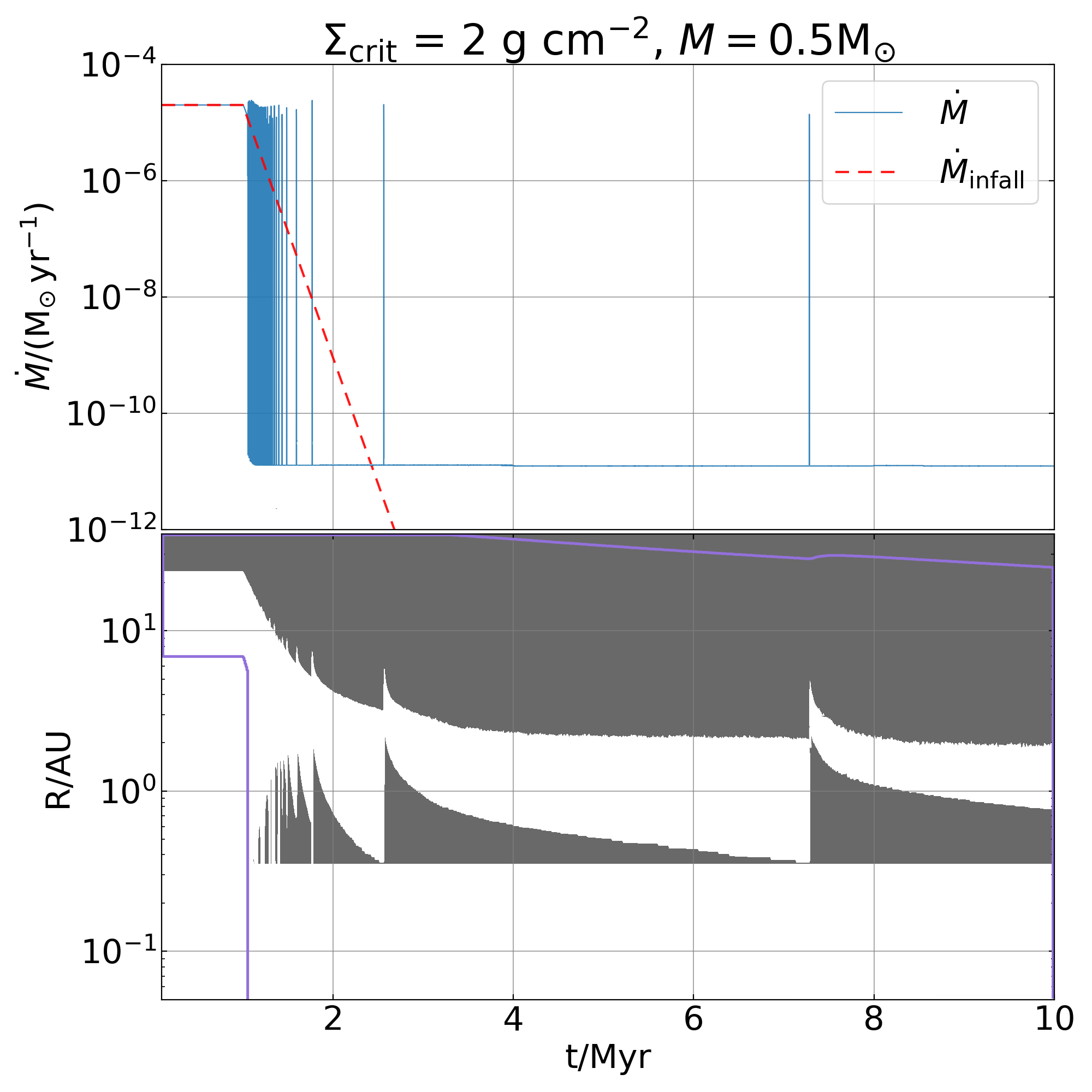}
  \end{subfigure}\hfill
  \begin{subfigure}[b]{0.48\textwidth}
    \includegraphics[width=\textwidth]{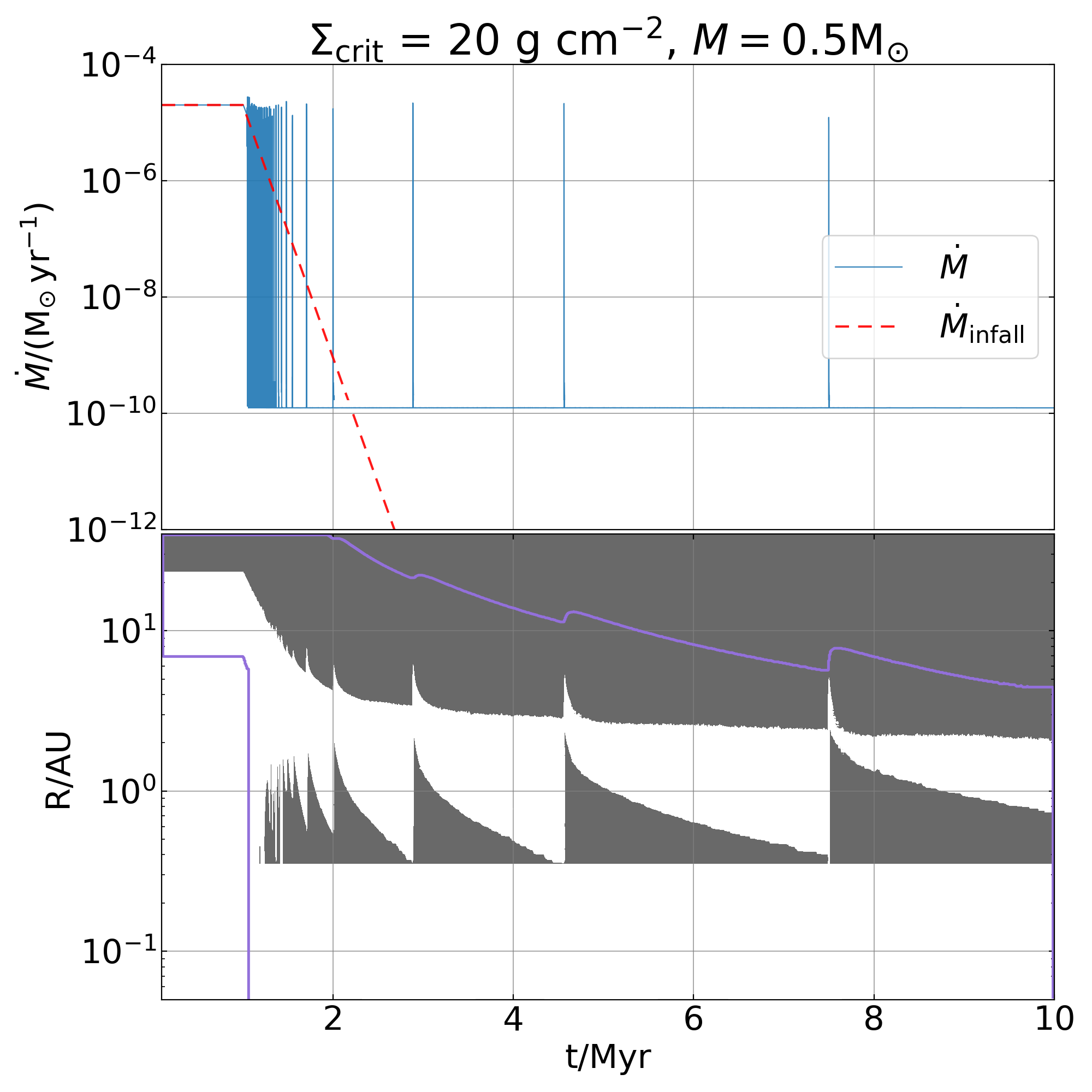}
  \end{subfigure}

  \vspace{0.6em} 
  \par 
  \begin{subfigure}[b]{0.48\textwidth}
    \centering
    \includegraphics[width=\textwidth]{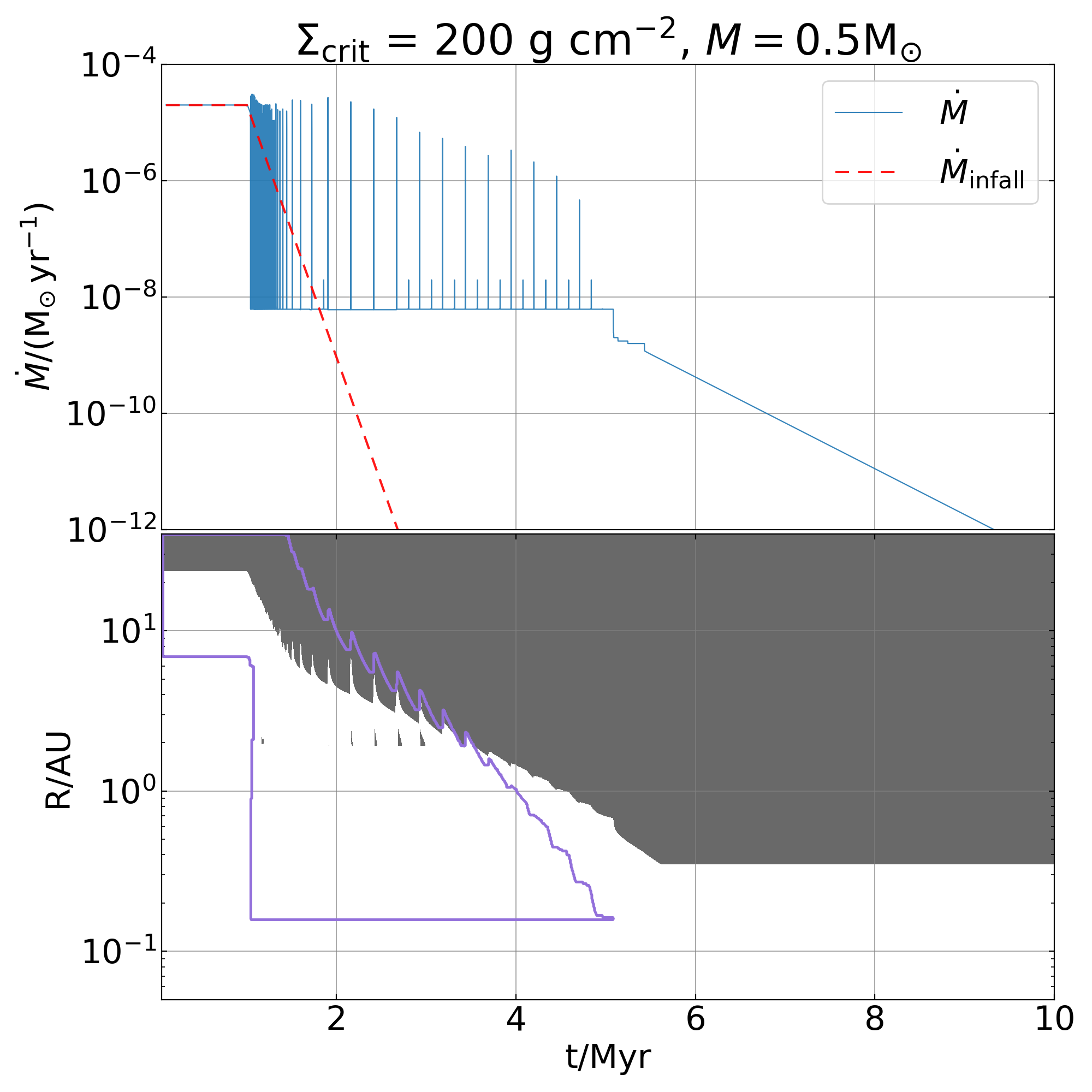}
  \end{subfigure}
    \caption{Same as Fig.~\ref{fig:graph1} except $M=0.5\,\rm M_\odot$.}
   \label{fig:graph05}
\end{figure*}

\begin{figure*}
  \centering
  \begin{subfigure}[b]{0.48\textwidth}
    \includegraphics[width=\textwidth]{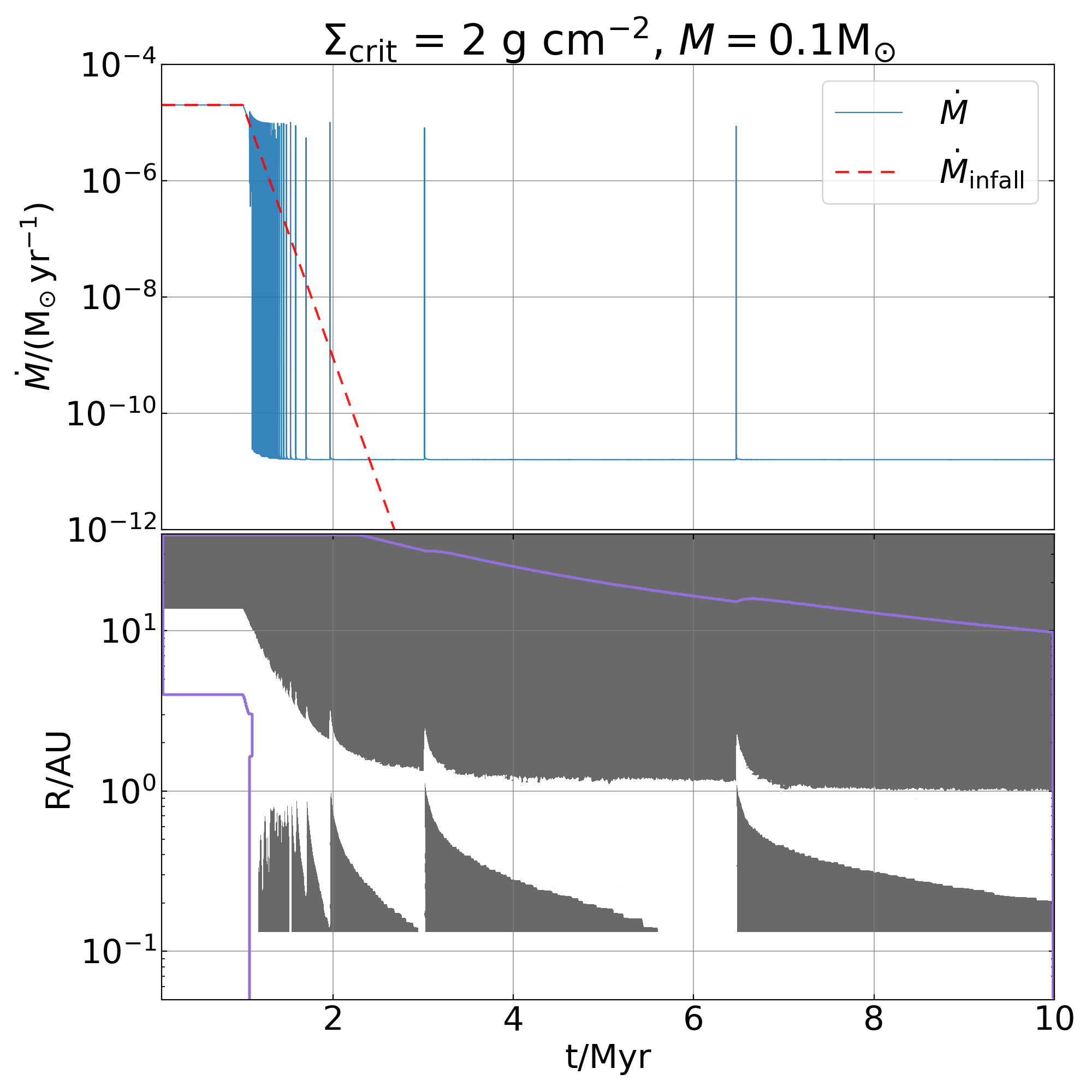}
  \end{subfigure}\hfill
  \begin{subfigure}[b]{0.48\textwidth}
    \includegraphics[width=\textwidth]{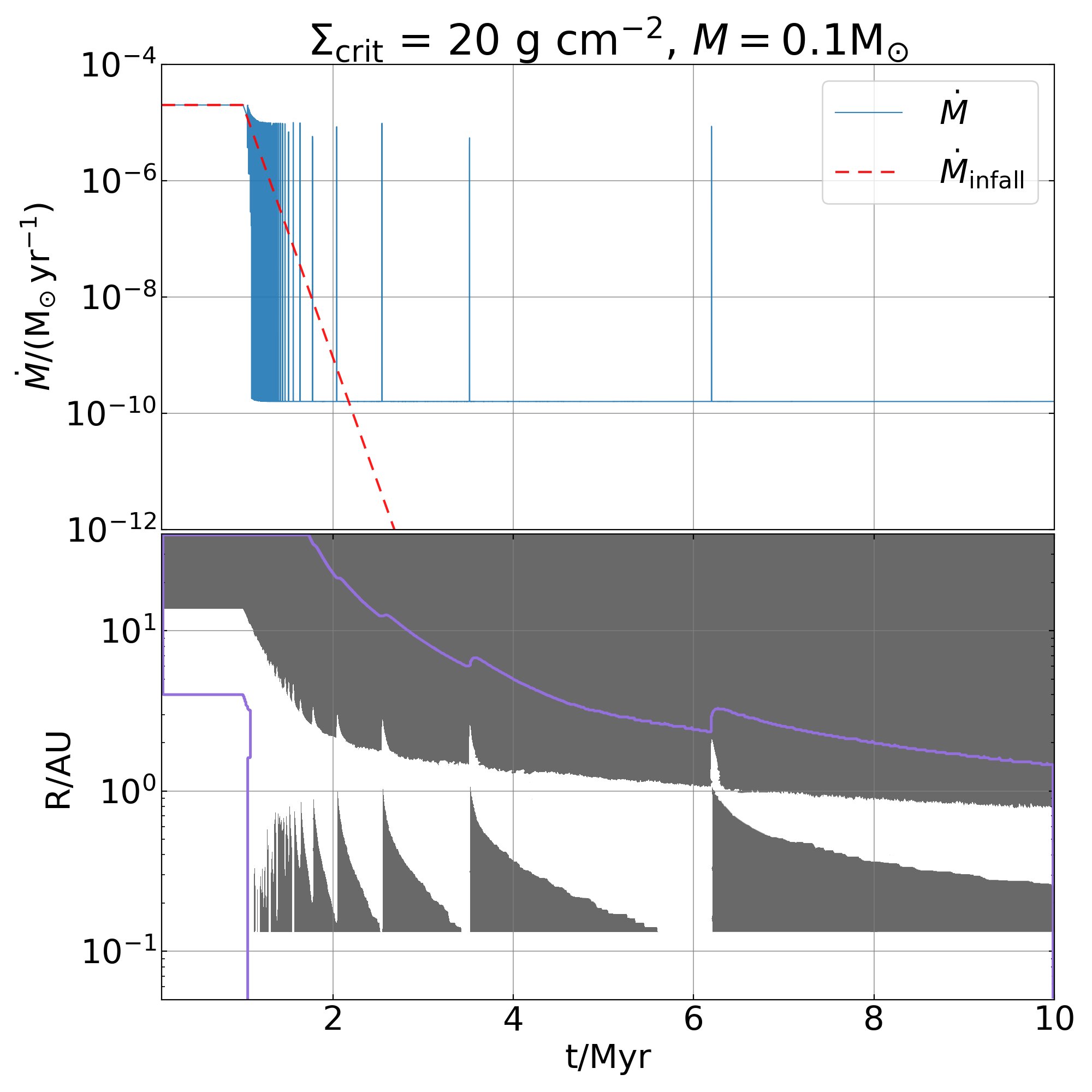}
  \end{subfigure}

  \vspace{0.6em}
  \par
  \begin{subfigure}[b]{0.48\textwidth}
    \centering
    \includegraphics[width=\textwidth]{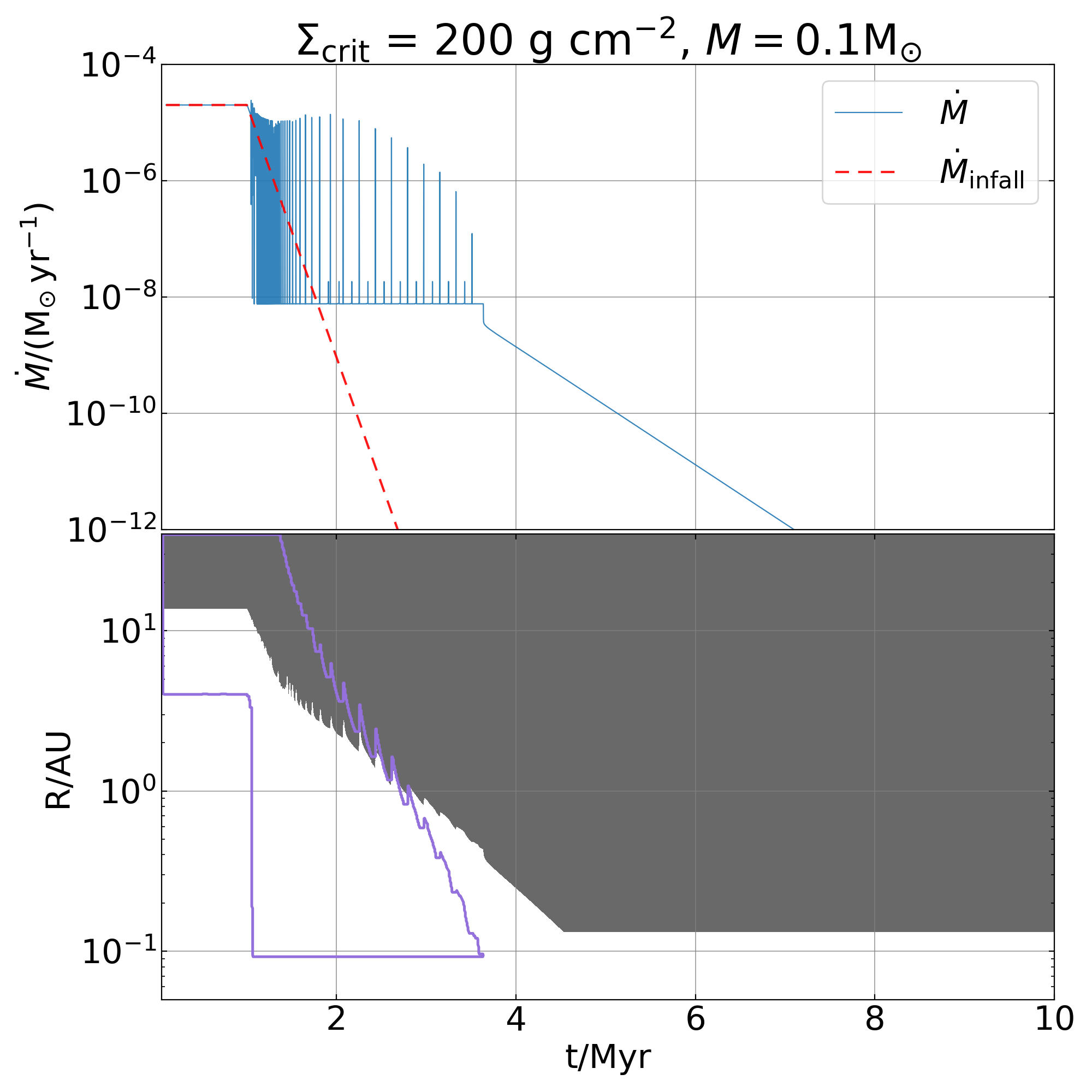}
  \end{subfigure}
    \caption{Same as Fig.~\ref{fig:graph1} except $M=0.1\,\rm M_\odot$.}
    \label{fig:graph01}
\end{figure*}

\begin{figure*}
  \centering
  \begin{subfigure}[b]{0.48\textwidth}
    \includegraphics[width=\textwidth]{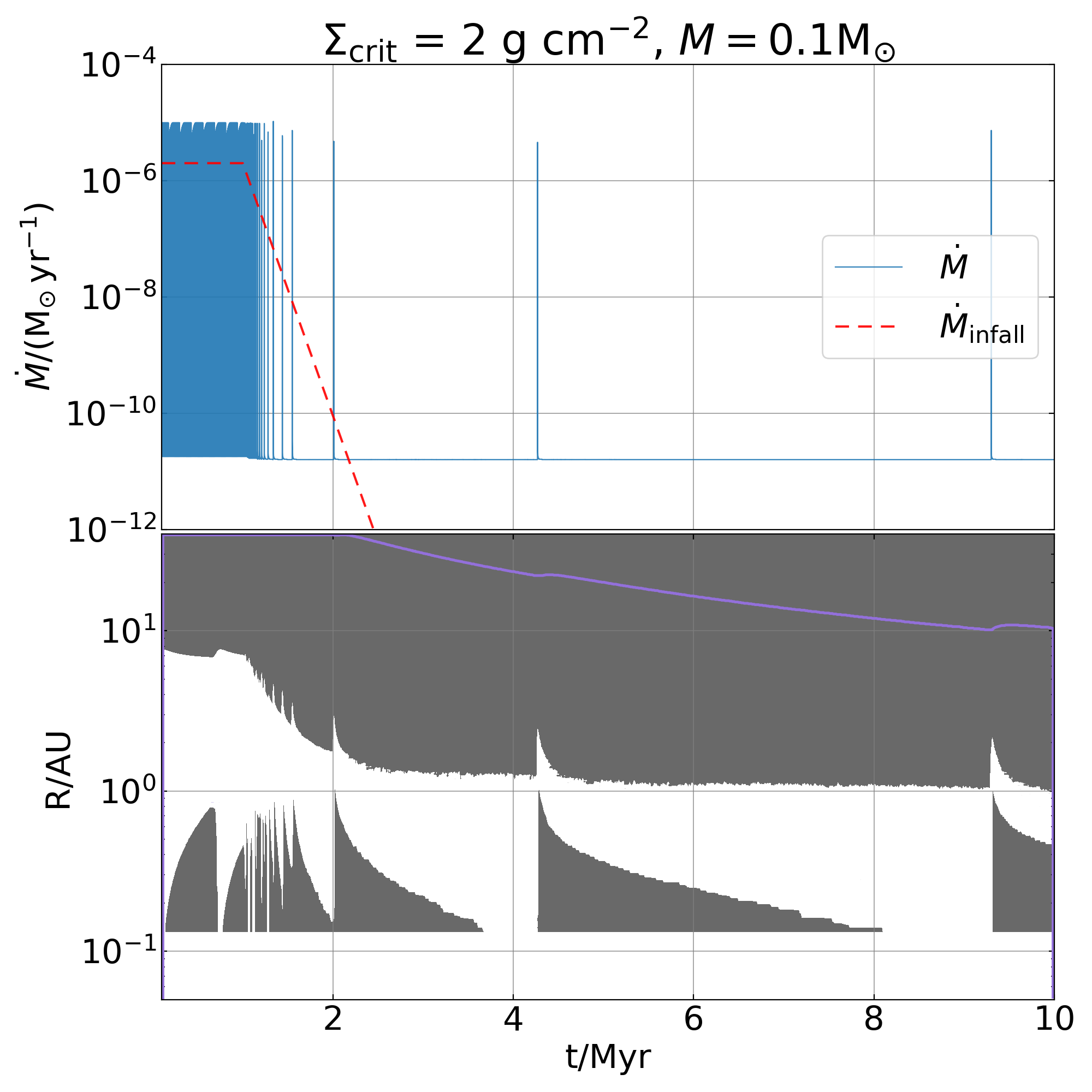}
  \end{subfigure}\hfill
  \begin{subfigure}[b]{0.48\textwidth}
    \includegraphics[width=\textwidth]{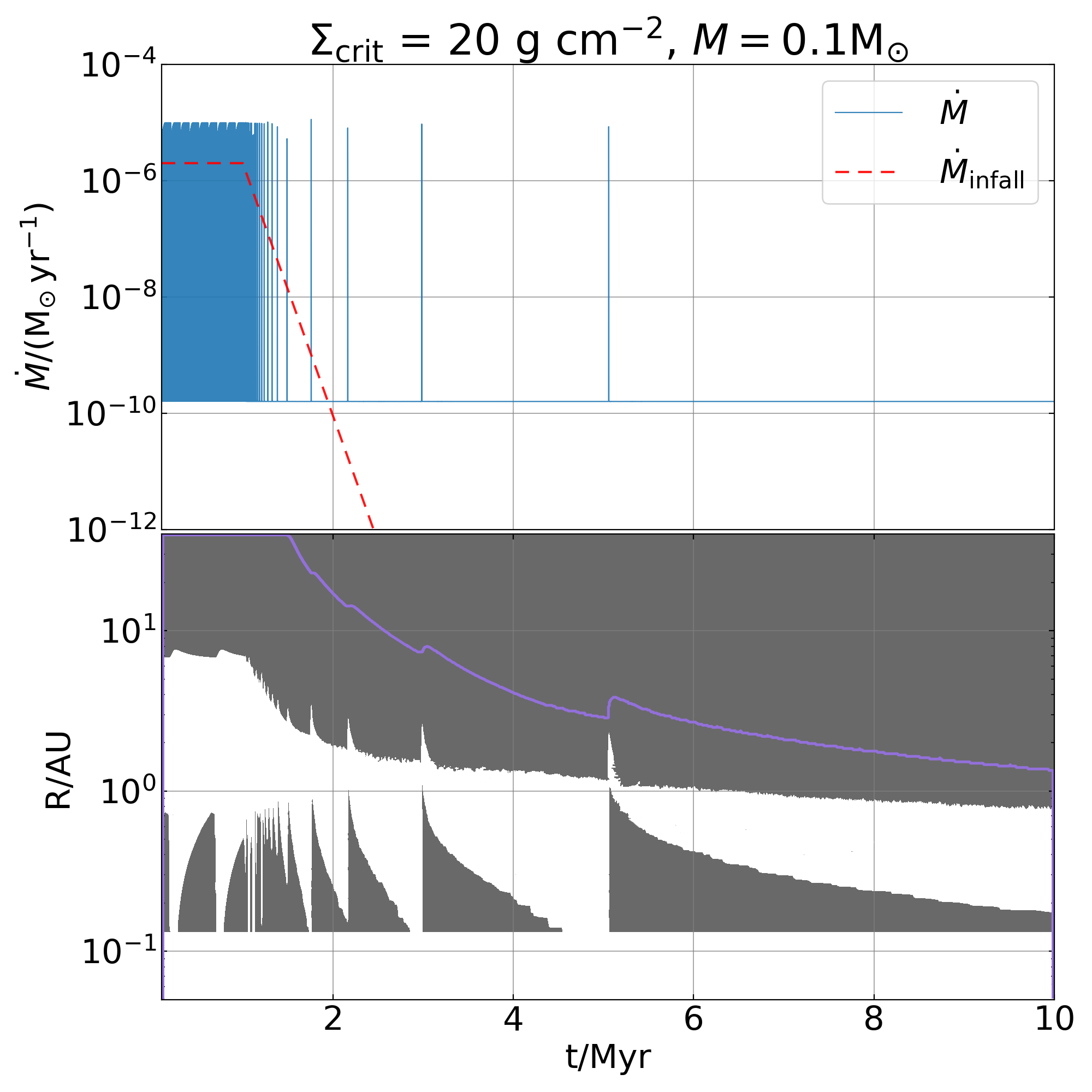}
  \end{subfigure}

  \vspace{0.6em}
  \par
  \begin{subfigure}[b]{0.48\textwidth}
    \centering
    \includegraphics[width=\textwidth]{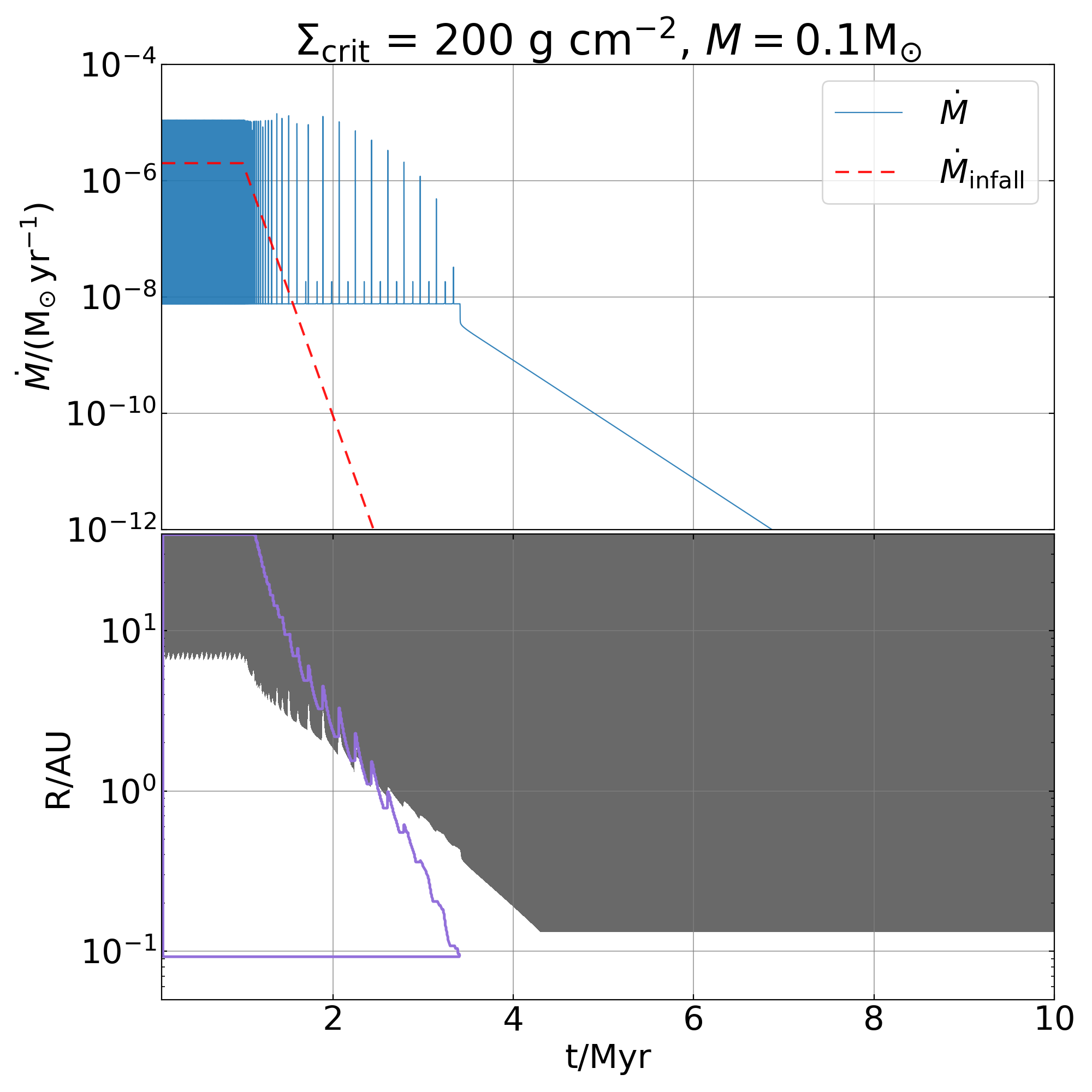}
  \end{subfigure}
      \caption{Same as Fig.~\ref{fig:graph1} except $M=0.1\,\rm M_\odot$ and the initial infall rate is $\dot M_{\rm i}=2\times 10^{-6}\,\rm M_\odot \, yr^{-1}$.}
    \label{fig:2e-6}
\end{figure*}

 We now numerically solve the time-dependent equations (\ref{eq:1}) and (\ref{eq:Tc}) together with equations (\ref{eq:Qplus}) and (\ref{eq:Qminus2}) to see how the inner icy region evolves in time. The simulations are performed for a time of $10 \,\rm Myr$. We employ a grid of 500 points uniformly distributed in $R^{1/2}$ for all cases. We have checked the impact of numerical resolution by repeating a subset of simulations with twice the number of grid points (1000 points) and found that the results remain  unchanged. The computational domain spans from the inner disc radius, $\rm R_{in}=5 \,\rm R_{\odot}$, to the outer disc radius, $\rm R_{out}=40\,$au. 
 
We keep the following parameters constant  $\mu = 2.3$, $\alpha_{\rm m} = 0.01$, $T_{\rm crit} = 800\,\rm K$, $Q_{\rm crit} = 2$. The full set of parameters adopted in these simulations is summarized in Table \ref{tab:params}. 
We impose a zero torque inner boundary condition $\Sigma(R_{\rm in}) = 0$ so that there is an inward flow of gas out of the grid and toward the central star. We apply a zero radial velocity condition at the outer boundary to prevent the flow from leaving through the outer boundary.

 \subsection{Infall accretion rate and initial conditions}
 
Material is added to the disc at a radius  $R_{\rm add}=35\,\rm au$. Initially, the infall accretion rate is given by $\dot{M}_{\rm i} = c_s^{3}/G$, where $c_{\rm s}$ is the sound speed in the cloud. For a temperature of $10\,\rm K$ \citep[e.g.][]{Wilson1997}, this is an infall rate of $\dot{M}_{\rm i} \approx 10^{-5}\, \rm   M_{\odot} \, yr^{-1}$. 
The simulations initially have a constant infall accretion rate for a time of $t_0=1\,\rm Myr$. This is long enough for the disc to reach either a
 fully turbulent steady-state accretion disc or a steady outburst limit cycle in the accretion rate. In a collapsing molecular cloud, the accretion rate onto the disc is not constant over time but declines exponentially \citep{Shu1977,Armitage2001}. Therefore, we choose the infall accretion rate 
\begin{equation} \label{eq:mdot}
 \dot{M}_{\rm infall} = \begin{cases}
  \dot{M_i} & {\rm for~~}  t<t_0\\
     \dot{M_i} e ^{\left( -\frac{(t-t_0)}{t_{\rm ff}} \right)} & {\rm for~~} t \ge t_0.
 \end{cases}
\end{equation}
The free-fall time scale given by
\begin{equation} \label{eq:tff}
 t_{\rm ff} = \left( \frac{3 \pi}{32G \rho_{\rm cloud}} \right)^{1/2},
\end{equation}
 where $\rm \rho_{\rm cloud} = 4.41 \times 10^{-19} \, g \, cm^{-3}$, is the density of the cloud that produces a free-fall time of $t_{\rm ff}=10^5\,$ yr \citep[e.g.][]{Armitage2001}.

 The initial surface density is given by
 \begin{equation} \label{eq:Sigmai}
\Sigma(R, t=0) = \Sigma_0 \left( \frac{R}{\rm AU} \right)^p,
\end{equation}
with $\Sigma_0=10^4\, \rm g\ cm^{-2}$ and $p= -1$.The temperature profile is given by 
\begin{equation} \label{eq:Tci}
T_{\rm c}(R, t=0) = T_0 \left( \frac{R}{\rm AU} \right)^q,
\end{equation}
with $T_0=10^4\, \rm K$ and $q= -0.75$, the initial temperature distribution is isothermal, given by $T_{\rm m}(R, t=0)=T_{\rm e}(R, t=0)=T_{\rm c}(R, t=0)$. These specific values for the exponents, $p=-1$ and $q=-0.75$, are chosen for consistency with theoretical models of protoplanetary discs, such as those described by \cite{Armitage2011}. However, the disc evolves to a steady state or a quasi-steady limit cycle within $\sim 10^4$ years, which is negligible compared to the disc lifetime. During this early phase, the disc is supplied by a constant infall accretion rate and therefore the simulation results are independent of these initial surface density and temperature conditions.\\

To demonstrate the evolution of the layered disc models, we conducted a total of twelve simulations. We consider the evolution of the disc with an initial infall rate of $\dot M_{\rm i}=2\times 10^{-5}\,\rm M_\odot \, yr^{-1}$ with three different stellar masses. Additionally, we show a case for the M-dwarf star with mass $M=0.1\,\rm M_\odot$ and an initial infall rate of $\dot M_{\rm i}=2\times 10^{-6}\,\rm M_\odot \, yr^{-1}$. For each combination of stellar mass and initial infall accretion rate, we consider three different critical surface densities. We describe these simulations in the next two subsections.

\subsection{Accretion rate onto the star}
\label{sec:acc}

The accretion rate through the disc is calculated with
\begin{equation} \label{eq:macc}
\dot{M} = 6 \pi R^{1/2}\frac{\partial}{\partial R} (\nu_{\rm m} \Sigma_{\rm m} R^{1/2} + \nu_{\rm g} \Sigma_{\rm g} R^{1/2})
\end{equation} 
\citep[e.g.][]{Gammie96}. The top row in each panel of Figs.~\ref{fig:graph1} ($M=1\,\rm M_\odot$), ~\ref{fig:graph05} ($M=0.5\,\rm M_\odot$), \ref{fig:graph01} ($M=0.1\,\rm M_\odot$), and~\ref{fig:2e-6} ($M=0.1\,\rm M_\odot$ and $\dot M_{\rm i}=2\times 10^{-6}\,\rm M_\odot \, yr^{-1}$) shows the accretion rate onto the central body as a function of time. The dashed red lines show the exponential decay of the infall accretion rate.

Figs.~\ref{fig:graph1} (1\,M$_\odot$), ~\ref{fig:graph05} (0.5\,M$_\odot$), and ~\ref{fig:graph01} (0.1\,M$_\odot$), show the disc evolution starting at $0.01\,\rm Myr$, during which the disc is in a fully turbulent steady state. The disc remains steady while the infall accretion rate is constant at $\dot M_{\rm infall}=2 \times 10^{-5}\,\rm M_\odot\, yr^{-1}$.  After $t=1\,\rm Myr$, the infall accretion rate drops exponentially in time, the temperature decreases, and a dead zone forms; the disc is no longer in a steady state. Then, the accretion proceeds in large outbursts with very small rates in between. Over time, there is not sufficient material left in the disc and outbursts occur less frequently as $\dot{M}_{\rm infall}$  decays. However, there can still be a dead zone in the disc structure to late times if $\Sigma_{\rm crit}$ is sufficiently small.

The value of the critical surface density, $\Sigma_{\rm crit}$, strongly regulates both the magnitude and frequency of the accretion outbursts. For a $M=1\,\rm M_{\odot}$ star, with $\Sigma_{\rm crit} = 2\, \rm g \,cm^{-2}$, $\Sigma_{\rm crit} = 20\, \rm g\, cm^{-2}$, and $\Sigma_{\rm crit} = 200\, \rm g\, cm^{-2}$ the highest accretion outbursts reach a magnitude above $\dot{M} \sim 10^{-5} \, \rm M_{\odot}\, yr^{-1}$, alternating with quiescent intervals where $\dot{M}$ falls to about $10^{-11}\, \rm M_{\odot}\, yr^{-1}$,  $10^{-10} \,\rm M_{\odot}\, yr^{-1}$ and $10^{-9} \,\rm M_{\odot}\, yr^{-1}$ respectively, according to the $\Sigma_{\rm crit}$ value. A similar trend is observed for  $0.5 \,\rm M_{\odot}$ and $0.1 \,\rm M_{\odot}$ stars, where lower values of $\Sigma_{\rm crit}$ correspond to reduced quiescent accretion rates. Therefore, the magnitude of $\dot{M}$ is more strongly affected by the value of the critical surface density than by the mass of the central body.

In addition to influencing the burst magnitude, $\Sigma_{\rm crit}$ also affects the frequency of outbursts. For smaller values of $\Sigma_{\rm crit}$, less material can flow through the MRI-active layers, allowing more mass to accumulate in the dead zone and resulting in more frequent outbursts. Conversely, for higher $\Sigma_{\rm crit}$, the increased mass flow through the surface layers leads to less accumulation in the dead zone and, therefore, less frequent outbursting activity. This demonstrates that both the amplitude and frequency of the accretion variability are regulated by $\Sigma_{\rm crit}$.

Additionally, in Fig.~\ref{fig:2e-6} we show a case where the disc does not begin from a fully-turbulent steady state. For $M=0.1\,\rm M_{\odot}$ and $\dot M_{\rm i}=2 
\times 10^{-6}\,\rm M_\odot\, yr^{-1}$  a limit cycle occurs in the first $1\,\rm Myr$ while the infall accretion is constant. During this time, the accretion outbursts are very frequent and they reach a magnitude of $\dot{M} \sim 10^{-5} \, \rm M_{\odot}\, yr^{-1}$, alternating with short quiescent intervals. At later times, after the infall accretion rate onto the disc decays, outbursts become less frequent due to insufficient material being transported through the disc to sustain new accretion events. The relationship between the critical surface density, $\Sigma_{\rm crit}$, the magnitude of $\dot{M}$, and the frequency of outbursts remains similar to the previously described cases. The influence of $\Sigma_{\rm crit}$ on accretion dynamics follows the same pattern, affecting not only the magnitude of the quiescent intervals but also regulating the frequency of outbursts as the disc evolves.

\subsection{Evolution of the snow line}
\label{sec:evolsl}

The bottom row of each panel in Figs.~\ref{fig:graph1} ($M=1\,\rm M_\odot$), ~\ref{fig:graph05} ($M=0.5\,\rm M_\odot$), \ref{fig:graph01} ($M=0.1\,\rm M_\odot$), and~\ref{fig:2e-6} ($M=0.1\,\rm M_\odot$ and $\dot M_{\rm i}=2\times 10^{-6}\,\rm M_\odot \, yr^{-1}$) show the evolution of the icy regions and the dead zone in the disc. The gray areas indicate icy regions where the mid-plane temperature $T_{\rm c}$ is less than the snow line temperature $T_{\rm snow }= 145 \, \rm K$. The purple lines show how the radial extent of the dead zone evolves. Irradiation sets the minimum radius that the inner edge of the dead zone can reach, defined by the condition $T_{\rm irr} = T_{\rm crit}$. For lower-mass stars, this minimum lies closer to the star. Consequently, the inner edge of the dead zone can evolve as the disc evolves, but it can never extend inward beyond this irradiation-defined limit. Furthermore, the extent of the dead zone increases as $\Sigma_{\rm crit}$ decreases. A larger $\Sigma_{\rm crit}$ results in thicker active layers that transport more material, leading to higher temperatures near the star and a less extended dead zone. The exponential decay of $\dot{M}_{\rm infall}$ after $t=1\,\rm Myr$ is reflected in the gradual inward expansion of the outermost icy region of the disc.

The innermost region of each disc is heated through irradiation and is not icy. In Figs.~\ref{fig:graph1} ($M=1\,\rm M_\odot$), ~\ref{fig:graph05} ($M=0.5\,\rm M_\odot$), and~\ref{fig:graph01} ($M=0.1\,\rm M_\odot$), we observe that when the disc has a steady state infall rate (until $t=1\,\rm Myr$), there is no inner icy region. The steady state disc has  $T_{\rm c}>T_{\rm snow}$ everywhere. Consequently, the only icy region is confined to the outermost part of the disc. However, as the infall accretion rate declines, a dead zone forms, and the disc is no longer in a steady state. Within this dead zone, self-gravity heats the disc, activating the MRI and significantly increasing the temperature, which leads to an outburst. 

During outburst phases, the inner icy region of the disc vanishes. After the outburst, once the disc cools, icy regions can reform within the dead zone, with their extent governed by the mass of the central star and the critical surface density $\Sigma_{\rm crit}$. Figs. ~\ref{fig:graph1}, ~\ref{fig:graph05}, and ~\ref{fig:graph01}  demonstrate that the inner icy regions are much smaller in radial extent and shorter-lived around a solar-mass star compared to an M-dwarf. Additionally, when the critical surface density is high, $\Sigma_{\rm crit} =200\,\rm g\, cm^{-2}$, the dead zone does not persist beyond a time of about $t=6\,\rm Myr$ and therefore the formation of the inner icy region is suppressed. The outermost disc regions remain cold and icy. Over time, as the infall accretion rate $\dot{M}_{\rm infall}$ declines exponentially, the outer snow line moves inward, expanding the outer icy region. However, there is no dead zone in this model, and that may be a requirement for planet formation to proceed. 

The lower rows in each panel of Fig.~\ref {fig:2e-6} illustrate the evolution of the snow line and the dead zone in the disc around the lowest-mass M-dwarf ($M=0.1\,\rm M_\odot$) with a lower initial infall rate. In this case, the disc does not begin from a fully turbulent steady state, but rather a quasi-steady outburst cycle. As a result, the inner icy region is present during the initially constant infall. Later, as the infall accretion rate declines, outbursts become less frequent and the inner and outer icy regions in the disc expand. Once again, when the critical surface density is high, the inner icy region remains small. The disc behavior is similar to the higher initial infall accretion rate at later times in the disc evolution. Therefore, our results are not very sensitive to our choice of the initial infall accretion rate.

\section{Conclusions}
\label{conc}
With 1D steady state disc solutions and time-dependent models, we have investigated how the protoplanetary disc structure depends upon the mass of the central star. The structure of a protoplanetary disc plays a crucial role in determining where solids can accumulate and grow, and where planets can form. A dead zone provides a quiescent region in the disc where solids can settle to the mid-plane and may be a requirement for planet formation to place. In addition, disc temperatures below the snow line temperature provide more solid material. We have explored how the in-situ formation of planets in an icy dead zone region depends upon the stellar mass.  

There may be two icy regions within a layered disc model, one far from the star and a second inner icy region within the dead zone.  We have shown that the radial extent of this inner icy region increases as the stellar mass decreases. The inner icy region around a solar-mass star is radially small and short-lived. However, around an M-dwarf, the icy region oscillates in size between accretion outbursts in the region $0.1-1\,\rm au$ and persists to the end of the disc lifetime. 

Additionally, we have shown that the icy regions are sensitive to the critical surface density in the surface layers that are MRI-active. When the critical surface density is high, $\Sigma_{\rm crit} = 200\,\rm g\, cm^{-2}$, the disc temperature is sufficiently high that there is no inner icy region even around an M-dwarf.  In this case, most of the low-density disc can be icy at radii $R\gtrsim 0.1\,\rm au$ since the snow line is determined only by the irradiation temperature. Similarly, the icy region around the solar-mass star is at radii $R \gtrsim 1.4\,\rm au$ for this large critical surface density. While the icy region may extend close to the star, the lack of a dead zone may suppress planet formation. However, for $\Sigma_{\rm crit} \lesssim 20\,\rm g\, cm^{-2}$, the inner icy region in the dead zone may be long-lived around M-dwarfs. If planet formation is favored in dead zones, these extended icy regions around M-dwarfs may help explain the observed abundance of close-in super-Earths around M-dwarfs.

\subsection{Discussion and Limitations}

First, we note that the icy regions presented represent a conservative estimate since we chose a low value for the snow line temperature and a low value for the critical temperature. We also considered simulations with $T_{\rm snow}=170\,\rm K$ and do not find significant qualitative differences to our conclusions on the presence of the dead zone around low mass stars. 

As the star evolves toward the main sequence, its radius and effective temperature change significantly. These changes directly impact the irradiation flux $F_{\rm irr}$ and the corresponding irradiation temperature $T_{\rm irr}$. Initially, during the pre-main-sequence phase, a solar-mass star is larger. As it evolves onto the main sequence,  the star contracts to a radius of approximately $R_* \approx 1\,\rm R_{\odot}$, and its surface temperature rises to $T_* \approx 5498 \, \rm K$  \citep{Eker18}. These changes affect the irradiation temperature as
\begin{equation}
T_{\rm irr} \propto T_* R_*^{1/2}.
\end{equation}
The ratio of the irradiation temperature in the main sequence and pre-main sequence is $\sim 0.8$. 

A similar evolutionary trend is observed for M-dwarf stars. For an M-dwarf with mass $M=\rm 0.5\,M_{\odot}$ the stellar radius contracts to approximately $R_* \approx 0.46 \,\rm R_{\odot}$, and the temperature of the star rise to $T_* \approx 3800 \, \rm K$. For an M-dwarf with mass $M=\rm 0.1\,M_{\odot}$ the radius contracts to $R_* \approx 0.24 \,\rm R_{\odot}$, and the temperature increases to $T_* \approx 3105 \, \rm K$ \citep{Fraknoi16,Eker18, Sarmento21}. The corresponding irradiation temperature ratio is also $\sim 0.8$. 

This change is not substantial enough to significantly change our conclusions on the icy regions within the disk.

We also considered a simulation with a higher viscosity parameter $\alpha=0.1$. This leads to more efficient angular momentum transport and faster accretion onto the star. In this case, material in the dead zone accumulates more quickly, eventually reaching the critical temperature $T_{\rm crit}$ required to trigger an outburst. This outburst allows the disk to cool more rapidly, increasing the extent of the inner icy region.  We also considered a larger $T_{\rm crit}=1600\,\rm K$.  In this case, the icy regions are slightly reduced in size because it becomes more difficult to trigger an outburst, preventing the disk from cooling down. However, the large changes that we have considered for these parameters do not significantly alter the plots that we have presented.

Our model has several limitations. We assume a constant value for $\Sigma_{\rm crit}$, whereas in reality this value may vary with ionization sources and chemical composition. While we have identified regions within the disc where temperatures are low enough for ice to form, we have not modeled the actual formation process. Finally, the 1D approach does not capture vertical structure or localized instabilities that may occur in dead zones. These simplifications could influence the exact location of the snow line and the lifetime of icy regions, particularly during episodic accretion phases.

\subsection{Future Work}

Incorporating detailed dust and ice grain evolution \citep{Birnstiel12, Okuzumi12}, along with planetesimal formation processes \citep{Johansen11}, could provide deeper insights into how these icy regions influence the early stages of planet formation \citep[e.g.][]{Liu2019,Liu2020}. Dust plays a key role in the structure and evolution of protoplanetary discs, by dominating the absorption and scattering opacity in most regions \citep{Beckwith90, Beckwith20, Bouwman20} transporting volatiles both radially and vertically \citep{Cuzzi04,Ciesla06, Oberg16, Krijt18} providing a surface area to promote chemical reactions \citep{Kress01, Ruaud19}, furthermore, its properties are inherited by planetesimals \citep{Jansson14}.

Since most systems are thought to experience frequent outbursts throughout their evolution, as proposed by the episodic accretion scenario \citep{Dunham12, Audard14}, modeling the growth and evolution of dust presents a significant challenge due to complex couplings between transport processes, disc conditions, and micro-physical properties of the dust grains. \cite{Houge2023} developed a local coagulation model based on the super-particle approach \citep{Zsom08} to simulate locally the coagulation and fragmentation of icy dust particles, their simulations have demonstrated how FUor-type accretion outbursts can alter the collisional evolution of dust and ice in protoplanetary disc mid-planes, leading to changes in e.g., the ice distribution and maximum size that persist long after the outburst has faded.  \cite{Ros24} showed that stellar outbursts reset particle size distributions, with the disc cooling timescale influencing the resulting particle sizes. They also highlight the importance of heterogeneous nucleation, a key process in particle growth where water vapor condenses on existing silicate particles as the disc cools.  \cite{Wang25} found that the water cycle makes active discs more conducive to planetesimal formation than passive disks. The significant temperature dip caused by latent heat cooling manifests as an intensity dip in the dust continuum, presenting a new channel to identify the water snowline in outbursting systems. 

Ultimately, comparing these theoretical models with high-resolution observations of protoplanetary discs \citep{Birnstiel18, Ansdell16}, particularly around M-dwarfs, will be essential to validate the predicted links between icy regions and the occurrence rates of super-Earths. This validation hinges on observations at milliarcsecond-scale angular resolution, such as those achieved by the Atacama Large Millimeter/sub-millimeter Array (ALMA). This capability is crucial for spatially resolving the thermal structure of the inner disk, allowing the direct imaging of snow lines at the critical scale of a few astronomical units (AU) where these planets are expected to form \citep{Andrews18, Long18}.

\section*{Acknowledgements}

We thank an anonymous referee for a careful and detailed report that significantly improved the paper. 

\section*{DATA AVAILABILITY}
\label{da}
The data underlying this article will be shared on reasonable request to the corresponding author.


\bibliographystyle{mnras}
\bibliography{references} 


\bsp	
\label{lastpage}
\end{document}